# A NEW FORM OF BANKING – CONCEPT AND MATHEMATICAL MODEL OF VENTURE BANKING

**Abstract**

This theoretical model contains concept, equations, and graphical results for venture-banking. A system of 27 equations describes the behavior of the venture-bank and underwriter system, allowing phase-space type graphs that show where profits and losses occur. These results confirm and expand those obtained from the original spreadsheet based model. An example investment in a castle that is worth half of the amount invested in it is provided to clarify concept.

This model requires that all investments are in enterprises that create new utility value. The utility value created, assessed by market pricing, is what the new money is based upon. Out of this the venture-bank and underwriter are paid. The model presented chooses parameters that ensure that the venture-bank experiences losses before the underwriter does. Parameters are: EDCS Premium, 0.05; Clawback lien fraction, 0.77; Clawback bonds and equity futures discount, 1.5·(USA 12 month LIBOR); Range of clawback bonds sold, 0 to 100%; Range of equity futures sold 0 to 70%.

**Correspondence:** Brian P. Hanley, Butterfly Sciences, California, USA

**Email:** brian.hanley@ieee.org

# 1 PREAMBLE

This working paper describes a theoretical system that improves the earnings of venture capital. It assumes two features that must be accepted by regulators. First, that when a loan is made to fund an at-risk real-economy investment (e.g. a standard venture capital convertible note in a real-economy enterprise) that the equity received for that loan principal is not necessarily subject to cancellation per double-entry bookkeeping and is realized as part of leveraged profits because this asset is a real one. There is ample justification for doing this (Hanley 2020). Second, that when an at-risk loan investment from the venture-bank does not return the principal amount, that this is simply forgiven. This has been done by central banks, and I have been told (although I do not have the citation) that this has been done at commercial banks in some instances. Again, there is logical justification for this, as the money that was created by the loan is only owed to the bank that originated it, by the bank that originated it.

Thus, there are two more possible models that could be developed, one which required full cancellation, or some fraction thereof, and another which did not forgive at-risk loans or some fraction thereof.

I have had one discussion with regulators who did not see inherent reason for objecting to these features in the case of truly at-risk investments in real-economy ventures. Most systemic risk occurs when the investments are in finance, and finance of finance to drive exponential bubbles. This has already been shown to be a problem (Hanley, 2012).

# 2 INTRODUCTION

The common conception of banking that many people have is that banks take deposits and then lend them out afterward. However, this not an accurate description today. In the modern world, banks do not lend money out that they collected as deposits. Instead, banks create deposits by lending, which is how most money is created today (McLeay, 2014). This fact must be understood in order to properly grasp how venture-banking works. In ordinary banking, reserves and deposits are found after the fact. While ordinary banking could, in theory, do what venture-banking does, without the requirement of loan coverage, this has proven problematic, in part due to the inherent moral hazard in banking. Requiring equity default clawback swap (EDCS) coverage sets up an adversarial relationship to put a brake on bad behavior. In addition, EDCS venture-banking, for the first time, matches the term of the investment to the money supplied to back it, and removes the requirement for deposits entirely. Taking advantage of money creation through loans, a venture can be profitable even if it does not return the loan amount in full. To illustrate how venture-banking works, I will start with a short story of building a castle using venture-banking where the castle is sold for half the cost to build it.

## 2.1 A VENTURE-BANK BUILDS A CASTLE – A TOY MODEL

A venture-bank writes a loan of $1 million to build a castle. This loan becomes an asset of the bank. The terms of the loan are that all equity in the castle is the property of the bank. When a bank writes a loan, it automatically creates new money sufficient to cover the loan (Gardner, 2006). However, regulators want to see that the bank has the funds in hand, which is a challenge of the

banking industry although it is not always strictly enforced. The issue is that the way the banking industry developed, there are multiple banks, and a loan may not be deposited into the bank that makes it, and as it is spent, it will end up at different banks. What this means is that some banks have a surplus of funds at times, and a deficit at others. Mechanisms like LIBOR provide for overnight borrowing to cover needs. But this method bypasses that, using a method that inherently matches the loanable capital term to the long-term needs of investors.

In this method, the loan is made and simultaneously an EDCS is purchased to cover the loan. That EDCS contract agrees to pay off any deficit should the castle not be worth the amount of the loan investment. In this case we will assume 5 years of coverage. The underwriter is paid 5% of the $1MM per year for 5 years, and at the end, will receive half of the equity in the castle if the castle investment has a return of 1 or better. The underwriter otherwise pays face value of the EDCS contract and receives all of the equity in the investment.

The underwriter receives $250K by the end of the 5 year term. The bank borrows the $250K to cover this loan at 2% interest. That works out to a $276K cost over 5 years.

5 years later, the castle is built. But, there is a problem.

The castle is valued at only $500K at completion, half the amount of the loan. So, the underwriter pays $1MM to the bank, and receives the castle asset, which is valued at $500K. Simultaneous with the EDCS payout, a lien for 77% of the payout is put on the bank, totalling $770,000. A pending credit for the castle asset value is active on the lien, but the underwriter has up to 180 days to accept or reject the asset valuation.

This $1 million payment retires the loan for the bank.

The underwriter has 180 days to accept the assessed asset value, deny the assessment, or find violations of the EDCS contract. We will accept the $500K valuation, assuming that the underwriter has not found any fraud, deception, or other evidence of violations of the contract. That official acceptance means that the base amount for the clawback calculation is cut from $1MM to $500K.

So we revise the clawback lien to 77% of $500K, which is $385,000.

Both parties now close out their transactions. The bank pays off the $385K clawback.

The bank's books:

| | | |
|---|---|---|
| | $1,000,000 | of EDCS payment |
| | - $ 276,000 | to pay off the total cost of the EDCS payments line of credit |
| | - $ 385,000 | to pay off the clawback |
| Total | $ 339,000 | ← Retained earnings on the $1MM loan and build. |

The underwriter's books have:

| | | |
|---|---|---|
| | $ 250,000 | in EDCS payments |
| | - $1,000,000 | payout on EDCS contract |
| | $ 500,000 | accepted assessed value of the castle equity |
| | $ 385,000 | clawback payments |
| Total | $ 135,000 | ← Net profit on cash flow investment of net $500,000 |

## 2.2 CASTLE EXAMPLE BOOKEEPING

Table 1: Venture-bank and underwriter. Loan asset shown in suspense with braces {}. created asset credit shown in brackets []. Retirement of an asset or lien shown with strikeout ~~9999~~. EDCS asset credit is cancelled when EDCS contract ends.

| Venture Bank | | | | Underwriter | |
|---|---|---|---|---|---|
| **Credit/Debit** | **Assets** | **Liabilities** | | **Credit/Debit** | **Running total** |
| | {$1,000,000} | | Investment loan | | |
| [$1,000,000] | | -$250,000 | DIN premiums (5 years) | $250,000 | $250,000 |
| | | -$26,000 | DIN premiums loan interest | | |
| $500,000 | | | Castle valuation | | |
| -$500,000 | | | Transfer castle. Assessment pending | $500,000 | *$750,000* |
| $1,000,000 | ~~{$1,000,000}~~ | | DIN payout | -$1,000,000 | -$250,000 |
| | | ~~-$770,000~~ | Clawback lien pending | | |
| | | -$385,000 | Clawback lien, castle equity accepted | | -$250,000 |
| -$276,000 | | | DIN loan & interest paid | | |
| -$385,000 | | | Clawback paid | $385,000 | $135,000 |
| **$339,000** | | | **Closed out investment** | **$135,000** | |

Table 2: Balancing of Castle example

| | **Balance** |
|---|---|
| $339,000 | Venture bank gain |
| $135,000 | Underwriter gain |
| $26,000 | Interest paid externally |
| $500,000 | Value created |

The key factor that allows this to work, is that when the bank creates a loan on its own behalf, that is newly created money that didn't exist before. If the bank doesn't get all of it back, whatever it does get back is still profit. This is the difference between a loss and a loan write-down.

## 2.3 A CASTLE BUILDING VENTURE CAPITALIST INVESTING USING A LOAN FROM A BANK DOESN'T WORK

After stepping through building a castle by a venture bank using this method of investment loans, the reader can see that the figures work. But some have remained suspicious, because it looks like "getting something for nothing". One fiduciary asked, "Why does this work when it could be cheaper to get a loan from another bank and pay the interest on that?"

There is a very simple answer, which is that after borrowing $1MM and having a castle asset worth only $500,000, the VC is down by $500,000 on their own account, plus the accrued interest of $131,408, for a total loss of $631,408. That comes out of their pocket, not the limited partners money. If the VC had only used limited partner money as usual, they would register a loss for the limited partners, but continue to draw their 2% management fee.

To break even when borrowing money from a bank the VC needs to be able to pay back the interest on the loan. Assuming a 2.5% interest compounded yearly for 5 years with no payments, they will have a balance 1.1314 times their principle. This would put them in the 57$^{th}$ percentile among VC firms.

However, perhaps a VC could use the EDCS system, to cover a conventional bank loan? I created that scenario using venture banking, and as you probably expect, it doesn't work.

Let us assume a more conventional scenario where the venture capitalist (VC) borrows $1MM to build a castle. He intends to sell it for double what it costs him to build it. In this scenario, the

bank would normally have the castle the VC is building as collateral for the loan. To make this scenario work, we must assume that the VC convinces the bank that he will pay for an EDCS, and promise that it will pay off the loan principal.

Unlike the venture-bank, the VC must make interest only payments to the bank of $25,000 per year for 5 years, for a total of $125K. He also must pay the EDCS premiums. We assume that the VC borrows both using a line of credit.

When the VC is finished building, the market has crashed, and our VC realizes that at an assessed value of $500,000 he can't pay back the bank.

Both parties now close out their transactions.

The venture capitalist's books:

$1,000,000 of EDCS payment
- $1,000,000 payment to the bank to close out the VC's loan
- $ 131,408 to pay off the accrued interest on the balloon loan
- $ 276,000 to pay off the total cost of the EDCS payments line of credit
- $ 385,000 to pay off the clawback

Total - $ 792,408 ← Total amount that the VC is short.

Clearly, our VC cannot pay off all of his debts from this castle deal. If it is the only deal that the VC has, it means bankruptcy. If the VC has any assets not tied to this deal, then the underwriter gets them, because this is a derivative instrument and operates prior to bankruptcy court proceedings.

The underwriter's books have:

$ 250,000 in EDCS payments
- $1,000,000 payout on EDCS contract
$ 500,000 accepted assessed value of the castle equity
$ 385,000 clawback payments (These could only be paid from other seized assets.)

Total $ 135,000 ← *Possible* net profit on cash flow investment of net $500,000

How much would a VC need to make to break even when using an EDCS covered 5 year bank loan? Assuming a 2.5% annual interest rate:

- $1,000,000 payment to the bank to close out the VC's loan
- $ 131,408 to pay off the line of credit used to make the interest only loan payments
- $ 276,000 to pay off the total cost of the EDCS payments line of credit

Total - $1,407,408 ← VC needs to make 40.74% over 5 years to break even.
This would be the 72nd percentile among VC firms.

It should be clear from these scenarios, that this is not a desirable way to operate for a VC. At best, our VC has other deals, but we know from Kaufmann fund's data (figure 1) that the median fund makes no money. The bottom three VC quartiles make 50% or less on their investments over 10 years.

## 2.4 DISCUSSION OF THE CASTLE TOY MODEL VS. REAL OPERATION

The differences between this scenario and the full venture-banking model of investing below, are several.

A. The model assumes that there is a large portfolio of investments which are the pool against which the underwriter is paid, not just one – this is the dataset taken from Kaufmann;

B. The model assumes EDCS contract payouts are at 5 years. This means that the underwriter must carry the payout expense for another 5 years, until the investments that return breakeven or better exit at 10 years. This means that the net underwriter investment in the above example is $250K plus carrying cost of $250K for 5 years, before collecting the $385K clawback.

C. In the model, the clawback lien takes 5 years to be paid off and accrues interest. This should satisfy regulators that the EDCS payout is real, as the venture-bank has use of that money until it needs to be paid back, which is usually years later.

Note that if the underwriter sells clawback shares at a discount then its true invested cost can be minimal or zero. Clawbacks represent a kind of bond backed by the venture bank, and sold by the underwriter. There is some room for tweaking exactly how this model is executed, and my choices were defined to maximize stability of the underwriter's business while being fair to venture-banks which are taking the lion's share of the risk. Keep in mind that if the underwriter fails, then the venture-banking system is over, so optimizing for stability of underwriters is paramount.

The model assumes that EDCS policies are written such that the clawback lien is against the bank, not the specific investment, and also assumes that the bank has a very large portfolio set. The Kaufmann Fund dataset represents on the order of 5000+ individual investments. Another way of working this would be to have the clawback lien be against some pool of investments, most likely a rolling set. Putting the lien on the whole bank, or on a rolling portfolio of investments is necessary to prevent the underwriters being unable to collect their due on heavier losses. (e.g. investments that return $0.126 on the dollar or less.) I have assumed that bad investments will be dumped faster than good ones, and in the model 5 years is the minimum EDCS premium period. In my model, all bad investments are dumped at 5 years.

This method of banking prevents perverse incentives by the bankers to under-assess the equity because of two things. First, perverse incentive is prevented because the underwriter collects a clawback lien at close. Second, perverse incentive is prevented because the bank gets credit for the equity in its insured investment, but this includes a type of claims process by the underwriter on the back end while the clawback lien is not paid off.

*The key fact to attend to is that this bank is making a loan to itself, for its own interests. That loan it makes for the investment is creation of new money, and that means that any fraction of that newly created money that the bank can hold onto becomes profit.* When there is a deficit, this becomes a write-down, not a loss in the ordinary sense. As we see above, if a VC has to pay off the loan to someone else first, then it becomes an actual loss. So venture-banking with EDCS, while it appears at first glance to be an expensive form of money, is actually quite efficient at creating profitability. At present, all venture capital investments that lose money are real losses, not write-

downs.

## 3 CONCEPT AND WALK THROUGH OF VENTURE-BANKING SYSTEM

*Basic concepts:*

1. Rate of return for venture-banks is controlled by the base capital of the bank, the multiple of loans that can be made from the base capital and covered by EDCS contracts, and the weighted average conventional return on all the investments. Thus, the average return on the portfolio times the multiple of original capital for the bank (MOC) becomes the total bank return.
2. The underwriter's Equity Default Clawback Swap (EDCS) contracts have a premium payment, and at exit, the underwriter receives a percentage of the exit equity.
3. When an EDCS instrument is called and paid out by the underwriter, it is terminated with a lien on the venture-bank that rides until exit of a portfolio of investments, when the portfolio of other investments hits year 10, or the lien is paid off. This lien is normally payable out of a specifically named investment pool. However, if the venture-bank does not have enough to pay them off at the portfolio exit, or the venture bank completely fails, then this becomes a liability of the VBU as a whole and would be paid out of assessments on the other venture-banks in the VBU.
4. Equity value created is the money out of which venture-bank gains, underwriter gains, and interest paid on carrying costs cannot exceed the equity value created. In the real world, business overhead will also factor in. However, for the venture-bank side, the 2% fee structure for management covers overhead for the VC part of venture-bank operations.

*Walk through:*

There are two primary parties, the venture-bank and the underwriters of Equity Default Clawback Swaps (EDCS). The EDCSs are written to pay the venture-bank the cash value of the note, and transfer all equity assets to the underwriter.

The venture-bank processes its investments as zero interest loans-at-risk, and buys an EDCS for each investment loan. Typically, the EDCS would be equal to the investment loan amount. An EDCS is classed as a derivative instrument, which makes its exercise immediate, and it cannot be stopped by bankruptcy or court action. For this reason, the clawback lien was created as the last element of derivative exercise.

The EDCS instruments allow the venture-bank to move their investment loans from their suspense account back into capital that can be loaned out. If an investment does not return the face value of the loan, then the venture-bank calls the EDCS, turns over all equity assets of the investment to the underwriter, and is paid the face value of the EDCS. The assessed value of the investment equity is credited to the venture-bank EDCS policy, and the clawback base is lowered by the amount of the equity assessment. When the EDCS call closes, the underwriter receives a lien on the venture-bank for a large percentage of the face value of the payoff, which is a clawback lien (Hanley, 2017a). Here, the clawback is assumed to be 77% of the net cash payout.

The lien on the venture-bank may allow the underwriter transparency into the operations of the venture-bank as they choose to investigate whether any fraud has occurred. The lien is not required to be paid off until the portfolio of the firm's other investments exit, the venture-bank voluntarily decides to pay it off, or some number of years have passed, nominally 10 years from start of the EDCS contract.

The underwriter can sell up to 70% of the equity futures in their EDCS portfolios prior to exits (Hanley, 2017b). The underwriter must, however, declare prior to each year’s operations, what percentage of its EDCS equity futures it is going to sell from its total portfolio. This is to prevent unloading junk on the public and keeping back the best investments, because that will destroy the market. The underwriter can sell up to 100% of its clawback liens as a type of bond, and similarly, must sell the same fraction of all of its clawbacks. In this modeling, both EDCS contracts and clawback liens sell at a discount with perfect pricing. These sales can improve the underwriter's cash flow and has major effects on ROI. It also has potential to insulate them against losses, since an EDCS share will be similar to LEAPS futures contract shares that the public can purchase. (LEAPS are long-term futures options, typically 1 year .)

When the venture-bank exits an investment, the exit equity value retires the investment loan and the difference is made up by the underwriter. The loan is made by the venture-bank for itself, and the exit equity held by the bank is unchanged by retiring the loan.

On a break-even or better investment, after retiring the loan, the underwriter's share of the equity (here 50%) is transferred to them. When a portfolio of investments are exited, the clawback liens will be paid. When the loan is below break-even, 100% of equity is transferred to the underwriter.

## 3.1 VENTURE-BANK EARNINGS

On a per VC investment deal basis, the venture-bank final net earnings[1] are:

(TotalExit_Equity − Underwriter_Equity_Share)
+ Gross_EDCS_Payout (If there is one.)
− EDCS_premiums with cost to carry
− Clawback_Lien
==============================
Earnings_per_VC_deal

To get the venture-bank's earnings, sum the earnings per deal for the multiple of capital (MOC).

$$\sum_{1}^{MOC} Earnings\ per\ VC\ deal \qquad \textbf{(1)}$$

In this modeling, the assumption is that investments are distributed across the Kaufmann fund’s portfolio returns at the level of granularity of a fund.

## 3.2 UNDERWRITER EARNINGS

EDCS_Premium_Total
+ Underwriter_Equity_Gains (The equity share given to the underwriter.)
+ Clawback_Lien
- Underwriter_Payout (The net payout loss realized by the underwriter.)
- Cost_of_Money_of_Payout
==========================
Underwriter_Earnings_per_deal

To get the underwriter's ROI per deal, divide the Underwriter_Earnings_per_deal by the total cost of payouts in equation 2.

$$\frac{Underwriter\ Earnings\ per\ deal}{Underwriter\ Payout + Cost\ of\ Money\ of\ Payout} = Underwriter\ ROI\ per\ deal \qquad \textbf{(2)}$$

To get the underwriter’s ROI in total, divide the sum of all underwriter earnings per deal by the sum of the underwriter payouts and cost of clawback money.

# 4 SIMPLIFIED MATH WALK THROUGH

## *4.1 FINANCIAL INSTRUMENTS*

- Venture capital will make their investments as loans from their own bank.
- Underwriters will write EDCS contracts for those loans.

The fundamental features of are:

- A yearly premium for the life of the EDCS. (Here, 5%.)
- When the venture exits the underwriter receives a percentage of the equity value. (Here, 50%)

---

1 The net earnings on the deal for the purposes of bank operation do not include clawback cost. The money in the clawback is operating capital.

- When the policy is triggered and paid off, the underwriter receives all current equity, and liquidates it quickly. (Underwriters may choose different schedules.) In this accounting, the underwriter pays the net difference between whatever the value of the equity is and amount of the investment loan.
- When the policy is paid, the venture-bank is immediately encumbered with a lien (clawback lien) for some fraction of the policy (Hanley, 2017a). (Here, 77%.) This lien does not need to be cured until exit of an investment pool or a time period has passed.
- Under no circumstances may the owner/beneficiary of the EDCS be separated from the owner/beneficiary of the investment it covers. (e.g. it is bound to the investment loan and must always go with it if the loan is sold.)
- The underwriter may not terminate the policy prematurely.

## 4.2 ASSUMPTIONS

- A venture-bank can create up to 47 times the original capital using this mechanism (Hanley, 2017b). However, over time, since investments are made and exited, this 47X multiplier can be exceeded in practice.
- I make a simplifying assumption that each investment is equal to the original capital. (e.g. all investments are the same amount.) However, use of calculus here provides us with the continuous version. If not, this would require a fairly complex agent model, and data for that type of model has not been found yet.
- I make an assumption that each turn of the capital yields the Kauffman portfolio 20 year average over a 10 year term.
  - I will further assume that to get closer to the true internal rate of return of the Kauffman portfolio, I should divide 31% by 80%, because Kauffman only sees the net of the 2% + 20% carry. This should be conservative. See figure 1 for 2% and 20% VC structure.
  - If we set 80% of profits = 31% (e.g. the profit amount above return of capital), then the total VC profits were 31% / 80% = 38.75%. This gives us a corrected VC multiple of 1.3875. (The 2% per year management fee built in to Kauffman's data is ignored here, and can account for operating costs.)
  - In the more intensive mathematical model that follows, a more detailed correction of the Kauffman data, point by point, shows that when taking profits above break even, the corrected multiple is significantly higher, slightly above 1.55 (see Appendix Dataset 2).
- I will assume an EDCS premium of 5% per year, an exit equity fraction of 50%, and a clawback lien of 77%. These figures can change, and are set conservatively so that underwriters will have a strong business.
- The clawback lien is a fixed percentage. At 77%, this works to produce a smoothly increasing profits curve for the venture-bankers, and should discourage "jackpotting" by deliberately bankrupting investments.
- I make a simplifying assumption that 50% of the EDCSs will pay premiums for 5 years, and then the venture-bank will collect its payoffs. For the balance, 50% of the investments will pay premiums for 10 years and then exit. (For the Kauffman fund data, 48.9% of investments lost money, and 51.1% made money. For the calculus model that follows, losers vary from 100% to 0%.) Note that these figures are not the same as the figures for internal VC operations. Venture capital in general assumes losing on 8 out of 10 or more investments. However, here, we are using higher level data of VC fund performance.

## 4.3 VENTURE BANK'S WALK THROUGH:

EDCS premiums
At 5% per year, over 10 years, accounting for cost of money at 2%, is 54.75% of loan principal.
54.75% x 50% of investments = 27.375%
At 5% per year, over 5 years, cost is 26.02% of loan principal.
26.02% x 50% of investments = 13.01%
Total cost of EDCS over 10 years = 40.385% of average loan principal.

EDCS payouts in this model are paid out in year 5. Underwriters make the venture-bankers whole for any investment loan that comes in at less than principal. For the Kauffman figures, when we do the modeling math, payouts work out to 13.61% of the average investment loan.

At exit, the multiple of 1.3875 is the total equity created by the venture-bank for investment. It receives this amount, and retires the loan. Note that after this loan is retired, the equity generated remains with the bank. So the valuation doesn't drop to 0.3875, it remains valued at 1.3875 because the venture-bankers are paying the loan they made to themselves off with earnings to themselves.

Next, the venture-bank owes 50% of the equity to the underwriters for investments that exit positively. However, the underwriter has already taken all of the equity in investments that are less than 1. Consequently, we don't pay half of the obvious total. In this example, the positive exit equity the underwriter has claim to is 1.028089.
1.028089 x 50% = 0.5140445 (Underwriter equity)

1.3875 – 0.5140445 = 0.8734555
0.8734555 is the fraction that is left for the venture-bank.

The venture-bank then must pay off the accrued cost of the EDCS (assumed to be borrowed funds from some other source), which is 0.40385.
0.8734555 – 0.40385 = 0.4696055

Now let's pay off the clawback. Clawback takes 77% of the extra earnings from payments. The venture-bank received 13.61% as payouts. 13.61% x 77% = 10.4797% that the venture-bank has to pay back to the underwriter at exit.
0.4696055 - 0.104797 = 0.3648085

The venture-bank has a net average 0.3648085 on each turn of capital.

Now we decide how many turns of capital the bank will make. Multiply by the MOC to find the venture-bank's earnings multiple.

47 x 0.3648085 = 17.14
43 x 0.3648085 = 15.68
30 x 0.3648085 = 10.944
5 x 0.3648085 = 1.82
4 x 0.3648085 = 1.459
3 x 0.3648085 = 1.09

From the above figures, we can see that it only takes 4 turns to do better than the results of the Kauffman portfolio (corrected to 1.3875 here) that we are basing this model on. And, it only takes 3 multiples of capital to equal the median venture capital fund.

### 4.4 UNDERWRITER'S WALK THROUGH:

First we calculate the premium fraction for the half of EDCSs that are paid for the entire 10 years.
At 5% per year, over 10 years, premiums are, 50% of the loan principal.

50% x 50% of EDCSs = 25%

Next we calculate the premium fraction for the half of EDCSs that are paid for 5 years.
At 5% per year, over 5 years, cost is 25% of loan principal. 5 years is used here because after an investment goes bad, no more premiums will be paid on it.

25% x 50% of EDCSs = 12.5%

Total premiums collected over 10 years = 25% + 12.5% = 37.5% of investment loan principal.

Total net payouts on year 5 is 13.61% of loan principal. On this, premiums equal to 12.5% of loan principal will be collected, leaving a net, uncovered cost of doing business of 1.11% of loan principal after 5$^{th}$ year payouts.

At exit, underwriter receives 51.40445% of loan principal in equity. Percentages below are percentages of the loan principal covered by the EDCS.
Net earnings per investment loan insured:

| | |
|---|---|
| 37.5% | Premiums |
| - 13.61% | Payouts on EDCS |
| - 1.416% | Cost to carry payouts for 5 years |
| + 10.4797% | Clawback payoffs |
| + 51.404% | Equity payment from venture-bank. |
| ======== | |
| 84.38% | Earnings, relative to the venture-bank's investments. |

The underwriter's multiplier for an ROI calculation is:

Cost to carry payouts for 5 years. At a 2% annual interest rate, this cost is 15.61%

84.38% ÷ (13.61% + 1.416%) = 5.62

The ROI multiplier should be higher because in the real world by year 5, only a small fraction of the payouts will not be covered by the premiums collected. I will not show this in the mathematical model below, because it complicates matters, and at this stage being conservative is more important. To illustrate this, if we assume that 1.1% of the loan principal is not covered by premiums when it comes time to pay them off, then the underwriter's ROI is over 77X. However, this model does not have an overhead figure.

## 5 MATHEMATICAL MODEL

Computations and graphs for this venture-bank model performed using Maple™ (Maple™, 2018).

### 5.1 VENTURE CAPITAL (VC) VERSUS VENTURE-BANK (VB) RETURNS

Here I will use Venture Capital (VC) portfolio and Venture-Bank (VB) returns, and these are not interchangeable. I use them because the venture capital dataset that I have is for VC portfolio returns in the conventional manner as shown in figure 1, managed per figure 2. Each one of those figure 1 VC funds is itself a portfolio. When I specify VC returns, this is a placeholder for the return on the investment for a specific portfolio. However, there may be higher variability for an individual deal. VB returns can only be generated from a set of investments using the MOC multiplier.

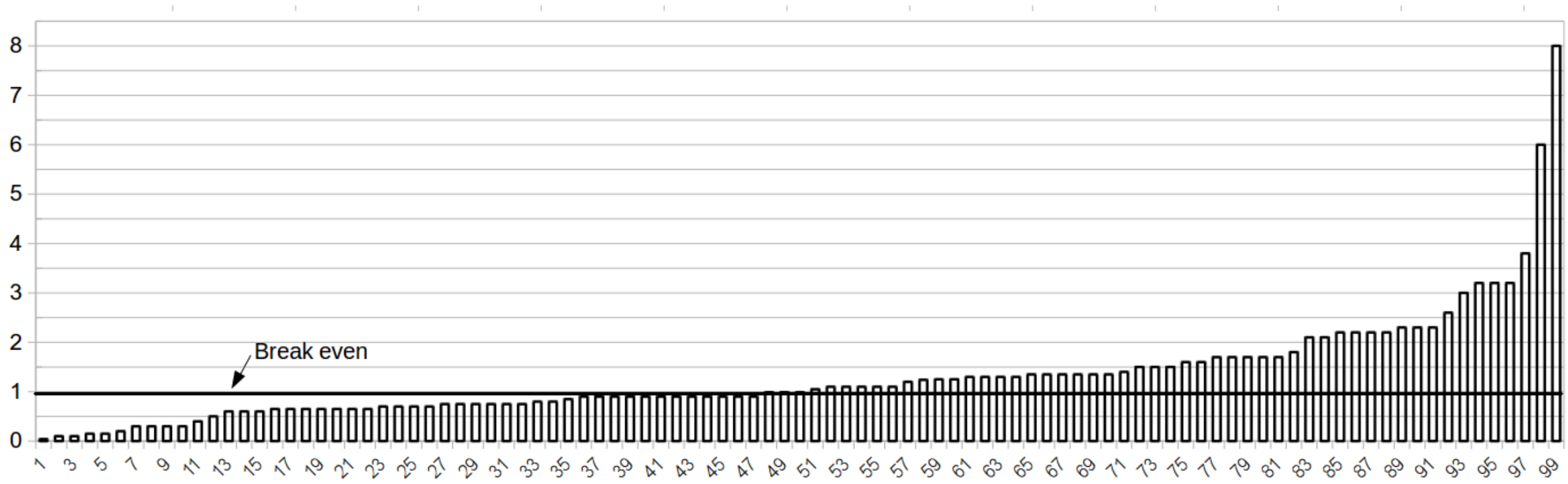

Figure 1: Kaufmann fund VC firm dataset. 99 funds performance over 20 years. (Mulcahy, 2012)

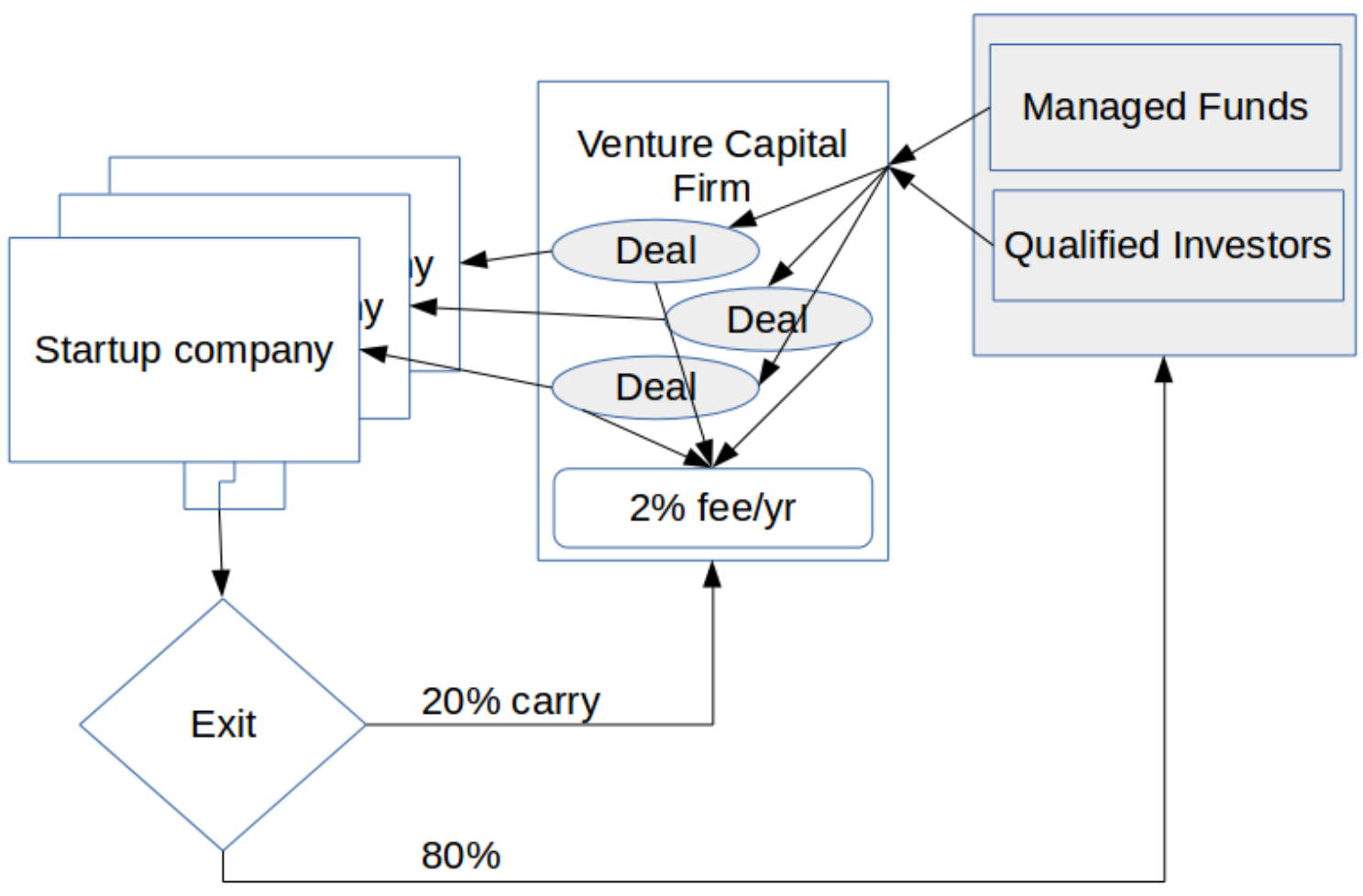

Figure 2: Venture capital primary model. In this modeling the 20% carry is taken at payout of the total fund when returns are greater than breakeven. (Used with permission from Hanley, 2017b)

## 5.2 ADJUSTMENT OF KAUFFMAN FUND DATASET

In figure 3, two plots are visible for the Kauffman dataset (Mulcahy, 2012). The gray-green dashed line shows the unmodified Kauffman data curve (see Appendix Dataset 1). However, here we want to look at the internal returns of the venture capital funds inside the Kauffman portfolio. As shown in figure 1, Mulcahy reports them as operating with a 2% per year management fee[2] and a 20% carry. The 20% carry is taken from profits, which means Kauffman sees only 80% of profits. So, for every Kauffman VC fund that returned greater than 1 after 10 years, that element of the portfolio was divided by 80%. This yields the red stepped line. (see Appendix Dataset 2) That finite element calculation significantly raised the total rate of return from 1.31 to 1.55 for the Kauffman

2 The 2% yearly fee is ignored. It can be considered accounting for internal operating costs.

portfolio of portfolios.

## 5.3 KAUFFMAN FUND DATASET CURVE FIT.

In this model, instead of an *x* axis, there is an *h* axis for the equations derived from the Kauffman data. There are two primary equations, a close fit, and an adequate fit. The close fit equation is, unfortunately, a 7th order exponential. This could be used in a software model easily enough, using numerical methods to find the intercept on the y axis. However, this kind of equation cannot be solved algebraically for *h* given a y value, so I did not use it in this model because it would take much more time to execute. This close fit 7th order exponential is provided for informational purposes, in case someone else has an interest in using it.

The equation used is a simpler exponential based on the natural logarithm base *e*. This simpler equation is amenable to algebraic solving for variables of the equation.

In figure 3, we see in the blue dashed line that when a value is subtracted from all points, the fitted curve can go below zero. However, I assume that in the real world, the valuation of an investment portfolio cannot go below zero. So it is necessary to substitute zero for the fitted curve result below the zero intercept. To do that, functions that invert the fitted curve function are needed. This is another reason why the exponential equation (eq. 6 & 7) was used.

### 5.3.1 Close fit is a 7th order equation regression

$$
\begin{aligned}
&-0.377+25.2906120002135\cdot h-291.398659319748\cdot h^2+1671.47290175033\cdot h^3\\
&-4939.37387988463\cdot h^4+7727.69689219724\cdot h^5-6072.99026870706\cdot h^6\\
&+1886.41489392694\cdot h^7 \qquad \textbf{(3)}
\end{aligned}
$$

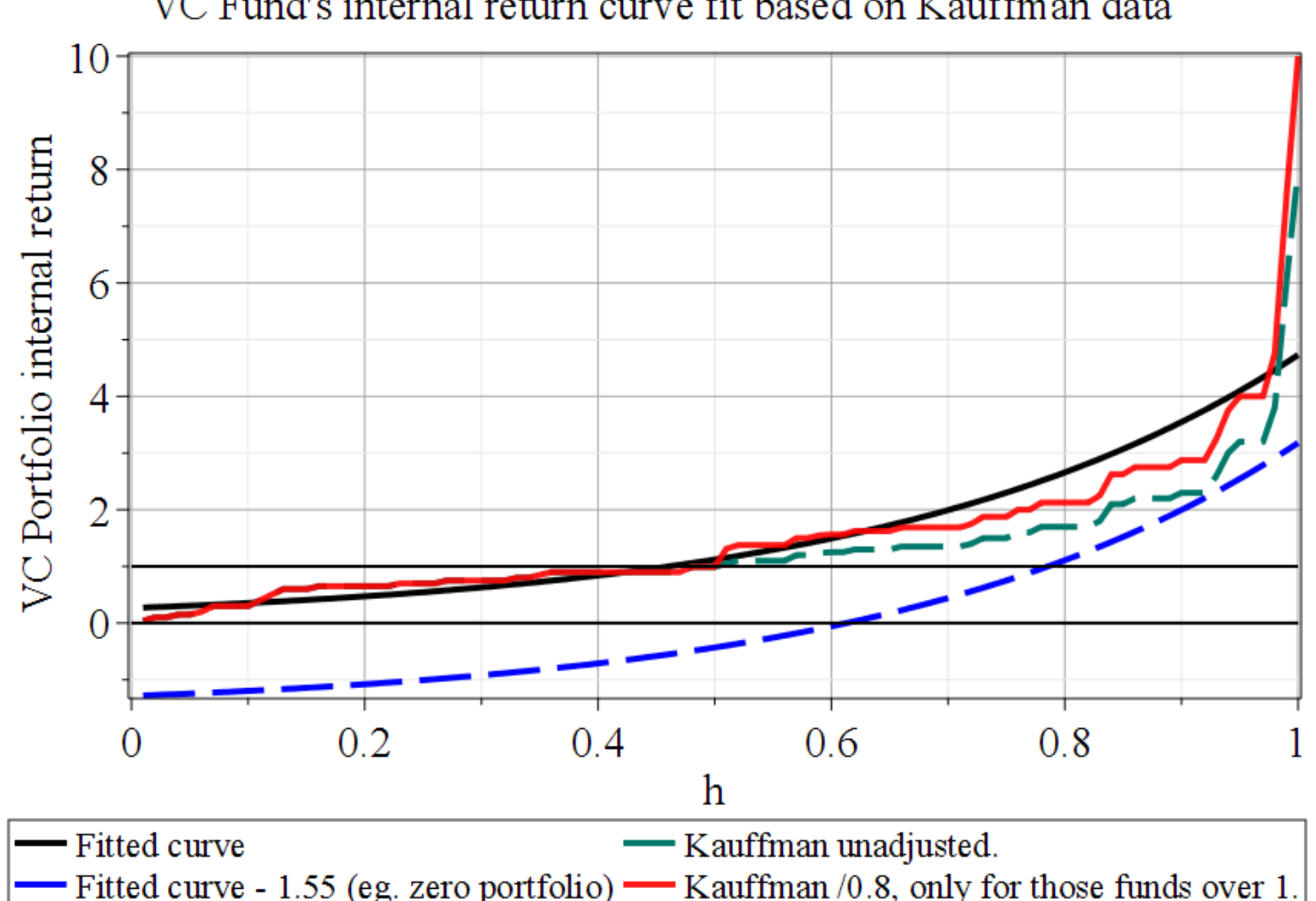

Figure 3: Kauffman data (gray-green dashed), Kauffman's adjusted graph (solid red), the exponential curve used in this model (black), and the net zero adjusted exponential curve used in this models. The blue dashed curve has a zero intercept of ~0.61, and a 1 intercept of ~0.78.

#### 5.3.1.1 *Integral of close 7^th order fit*

$$\int_{0}^{1.0} -0.377+25.2906120002135\cdot h-291.398659319748\cdot h^{2}+1671.47290175033\cdot h^{3} -4939.37387988463\cdot h^{4}+7727.69689219724\cdot h^{5}-6072.99026870706\cdot h^{6} +1886.41489392694\cdot h^{7}\,dh = 1.310174408 \qquad (4)$$

#### 5.3.1.2 *Indefinite integral of close 7^th order fit*

$$\int -0.377+25.2906120002135\cdot h-291.398659319748\cdot h^{2}+1671.47290175033\cdot h^{3} -4939.37387988463\cdot h^{4}+7727.69689219724\cdot h^{5}-6072.99026870706\cdot h^{6} +1886.41489392694\cdot h^{7}\,dh =$$

$$\mathbf{-0.377\,h + 12.645306\,h^{2} - 97.13288644\,h^{3} + 417.8682254\,h^{4} - 987.874776\,h^{5} + 1287.949482\,h^{6} - 867.5700384\,h^{7} + 235.8018617\,h^{8} \mid 0\,..1} \qquad (5)$$

### 5.3.2 Exponential regression, adequate fit – used in model

Figure 2 black line.

$$P+0.2655\cdot e^{2.88\cdot h} \qquad (6)$$

Figure 2 blue dashed line.

$$P-1.55+0.2655\cdot e^{2.88\cdot h} \qquad (7)$$

where: *P* is the average return of the venture capital (VC) portfolio.

#### *5.3.2.1 Integral of alternate exponential regression – used in model*

$$\int_0^1 P-1.55+0.2655\cdot e^{2.88\cdot h}dh \;=\; 0.000 \qquad (8)$$

where: $P$ is the average return of the venture capital (VC) portfolio.

Indefinite integral of alternate exponential with a portfolio factor to raise or lower VC portfolio return.

### 5.3.3 Solution of the *h* intercepts and boundaries

Alternate exponential solved for $h$ can give us the 1 (one) and 0 (zero) intercepts. To do this, create a variable that is the $h$ axis outcome, $y$, separate from the $h$. For both equations, set $h$ to 0.

$$y=P-1.55+0.2655\cdot e^{2.88\cdot h}\Rightarrow\frac{y-P+1.55}{0.2655}=e^{2.88\cdot h}\Rightarrow\ln\left(\frac{y-P+1.55}{0.2655}\right)=2.88\cdot h\Rightarrow\frac{\ln\left(\frac{y-P+1.55}{0.2655}\right)}{2.88}=h \qquad \textbf{(9)}$$

$$y=1\rightarrow\quad\frac{\ln\left(\frac{1-P+1.55}{0.2655}\right)}{2.88}\qquad\textbf{(10)}\qquad\qquad y=0\rightarrow\quad\frac{\ln\left(\frac{1.55-P}{0.2655}\right)}{2.88}\qquad\textbf{(11)}$$

Solve both of the above equations for $P$ when $h = 0$ and 1, to get the limits of the functions. Referring back to figure 2, we see that 0 and 1 are our limits.

$$\frac{\ln\left(\frac{2.55-P}{0.2655}\right)}{2.88}\rightarrow P=\quad 2.55-0.2655\cdot e^{2.88\cdot 0}\;=2.2845\quad=1\text{ intercept max.}$$

$$2.55-0.2655\cdot e^{2.88\cdot 1}\;=-2.179689529=1\text{ intercept min.}$$

$$\frac{\ln\left(\frac{1.55-P}{0.2655}\right)}{2.88}\rightarrow P=\quad 1.55-0.2655\cdot e^{2.88\cdot 0}\;=1.2845\rightarrow P=0\text{ intercept max.}$$

$$1.55-0.2655\cdot e^{2.88\cdot 1}\;=-3.179689529=0\text{ intercept min.}$$

where: $P$ is the average return of the venture capital (VC) portfolio.

These functions are named ProcIntercept1 and ProcInterceptZero respectively in the equations that follow. They have a parameter of $P$.

where: $P$ is the average return of the venture capital (VC) portfolio.

The graph verifying the results of these functions is in the appendix (See 9.3).

### 5.3.4 Revised 2 part equation for exponential fitted curve returns

$$\int_0^1 P-1.55+.2655\cdot e^{2.88\cdot h}dh-\int_0^{ProcInterceptZero(P)} P-1.55+.2655\cdot e^{2.88\cdot h}dh \qquad \textbf{(12)}$$

This equation gives us the net. This is the shaded blue area of figure 4.

where: $P$ is the average return of the venture capital (VC) portfolio; $e$ is number $e$

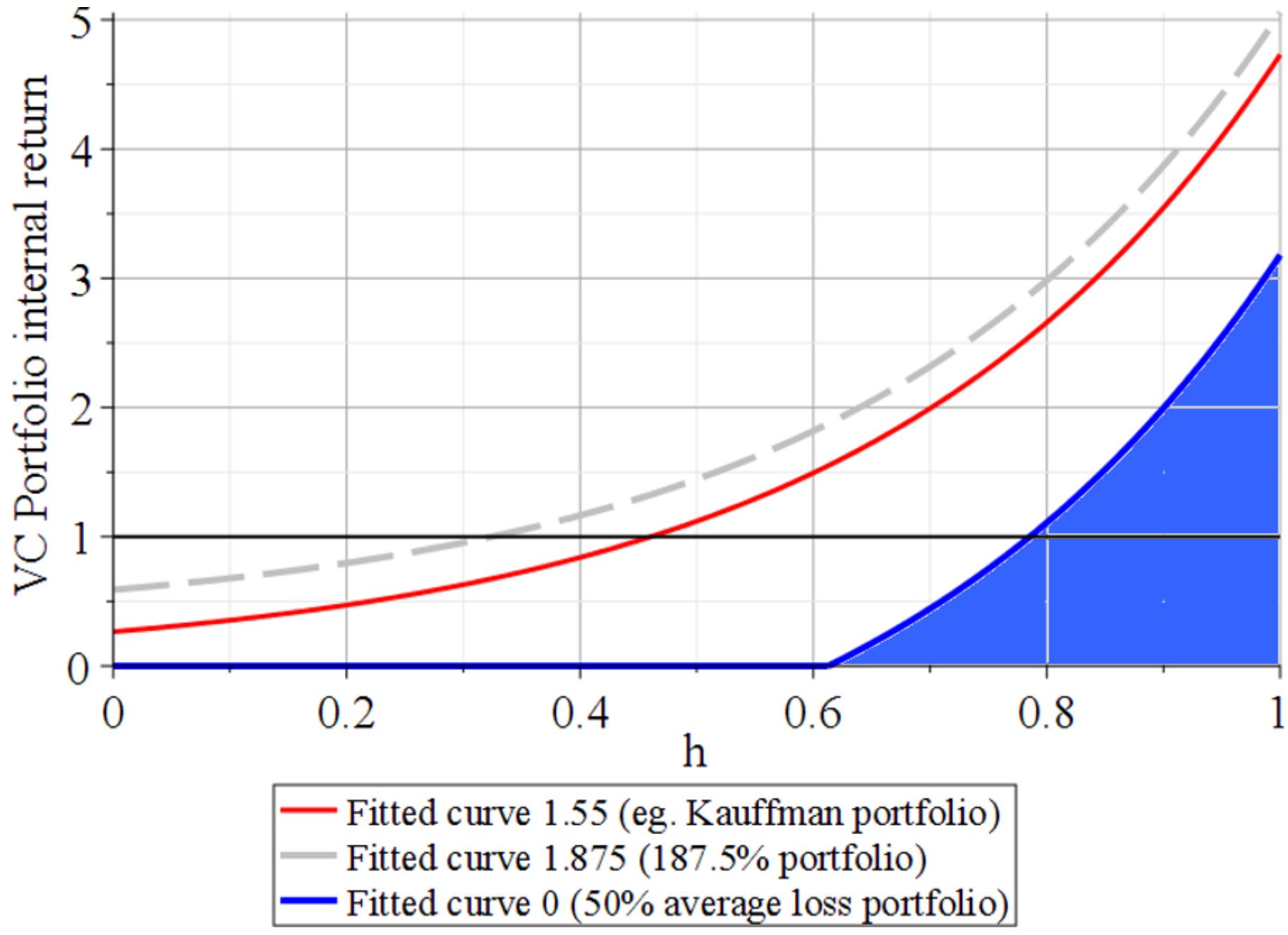

Figure 4: Sample curves for the chosen exponential. Red is the Kauffman portfolio adjusted to 1.55. Blue is a sample curve for a set of investments that go to zero. Light gray dashed line is the maximum probable large portfolio based on historical venture capital data.

## 5.4 EDCS PAYOUTS

The losses integral (eq. 13 & 14) gives us the total amount that the investment returns are below 1 for the region from 0 to the $y$ =1 intercept. These are the Venture-Bank losses, which are the same as the payouts. This formula assumes that the underwriter will accept the assessed valuation of the investment equity turned over by the venture-bank. This assessed valuation will lower the payout by the assessed valuation.

Note that a second reasonable scenario is that the underwriter takes over the equity and waits until some $n$ year hold time is complete. This latter requires somewhat different accounting. But then, the final exit equity value on the back end goes to the underwriter. However, I think that most underwriters would move to sell rapidly.

### 5.4.1 Definite form of losses integral

Venture-bank deal losses = Underwriter payouts

$$\int_{ProcInterceptZero(P)}^{ProcIntercept1(P)} 1-\left(P-1.55+0.2655\,(e)^{(2.88*h)}\right) dh \qquad \textbf{(13)}$$

When $P$ = 1.55, function = 0.2054307077

### 5.4.2 Definite form of losses integral for underwriters

$$(1+Intrsti)^{5} \cdot \int_{ProcInterceptZero(P)}^{ProcIntercept1(P)} 1-\left(P-1.55+0.2655\,(e)^{(2.88*h)}\right) dh \qquad \textbf{(14)}$$

where: $Intrsti$ is the interest rate charged for cost of money; $P$ is the average return of the venture capital (VC) portfolio; $e$ is number $e$

When $P$ = 1.55, function = 0.2279261069

## 5.5 CLAWBACK LIEN

The clawback lien occurs at the end simultaneously with the underwriter paying off the EDCS. It is a fraction of the net EDCS payout. (e.g. a fraction of the shortfall between the loan amount insured and the valuation of the equity accepted by the underwriter.) 77% appears to be an optimum. (See discussion below: Choosing a clawback lien fraction.)

### 5.5.1 Definite form of clawback integral

*Clawback* = 0.77 (77%)

$$Clawback \cdot \int_{ProcInterceptZero(P)}^{ProcIntercept1(P)} 1-\left(P-1.55+0.2655(e)^{(2.88*h)}\right) dh \qquad \textbf{(15)}$$

where: $P$ is the average return of the venture capital (VC) portfolio; $e$ is number $e$.
When $P$ = 1.55, function = 0.1581816449

### 5.5.2 Choosing a clawback lien fraction

First, a purpose of the clawback lien is to make it feasible to determine and have recourse on the back end if the bank is gaming/defrauding the underwriter. Since these are derivatives they are enforced immediately and, underwriters cannot implement a claims process prior to payment. The clawback allows a kind of back-end claims process after the fact should that be needed.

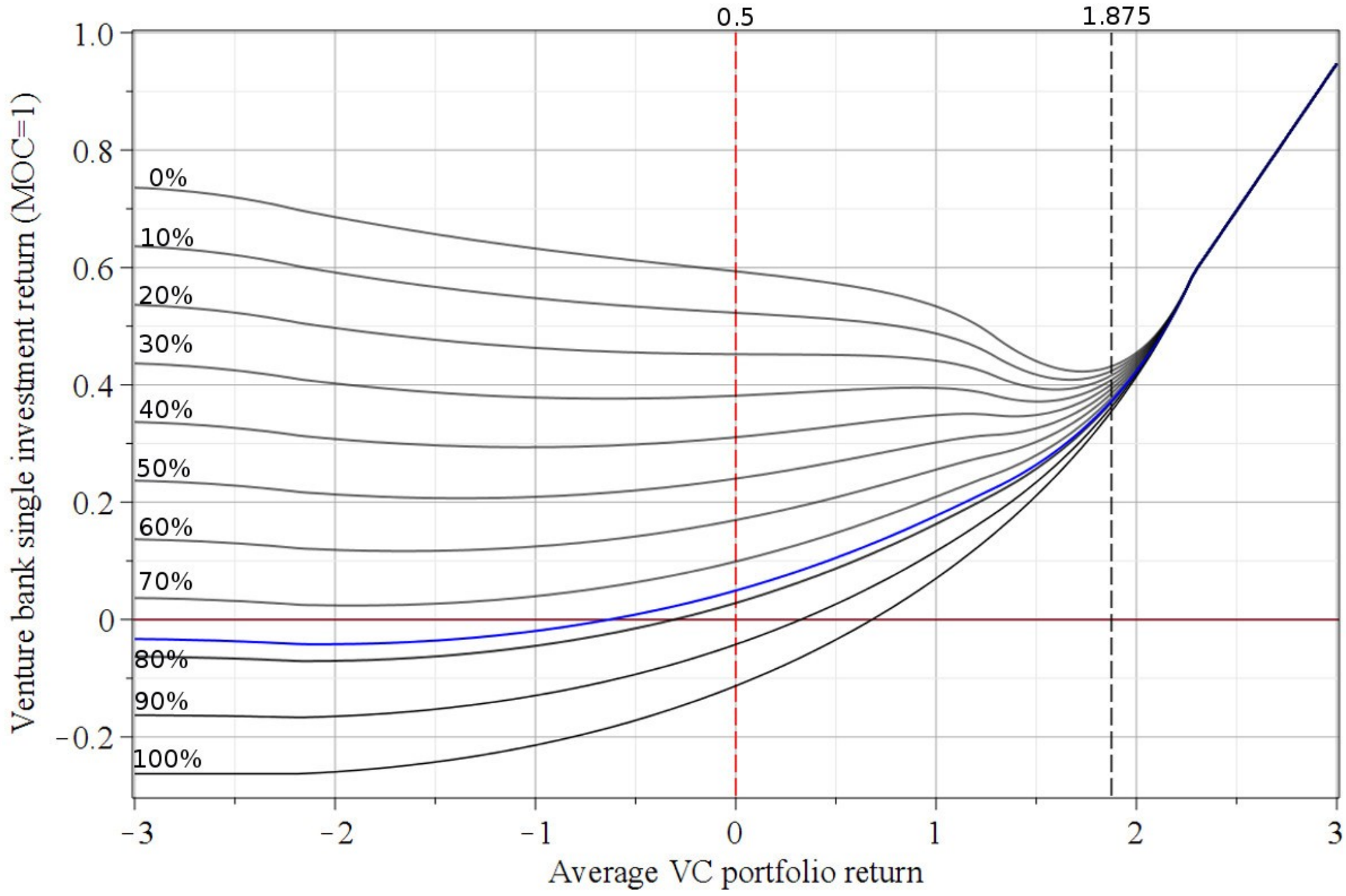

Figure 5: -3 VC portfolio adjustment corresponds to ~100% losses for a portfolio. This graph shows net venture-bank results for an MOC of 1 (no multiplier) with varying levels of clawback. Blue line is 77%. Most portfolios should fall within the 0.5 to 1.875 region. The most common good portfolios should fall in the bottom of the trough. Most fiduciaries would choose to make more money by lowering their VC portfolios if the clawback lien allows it to happen.

Second, the clawback lien is to ensure that there is a smoothly rising curve of profitability for venture-banks that receive payoffs. In figure 5, we see that without a clawback there is a strong motivation to commit fraud on the underwriter, and intentionally crash investments to make more money.

Fully 89% of the Kauffman VC funds returned 0.5 or better. Consequently I have chosen 0.5 (which is zero on the VC portfolio adjustment scale) as the lower end of normal operations that an underwriter should be dealing with. The normal high end is a bit below 1.875, as there is an example of a large pension portfolio without internal data that has returns of 1.5 . Divided by 80%, this is 1.875, which establishes a reasonable maximum.

In the selection graph of figure 4, we see a set of curves with different clawback values, as the VC portfolio return *(P)* factor varies. To pick a clawback fraction, decide what the lowest reasonable total portfolio return is. Then pick a value that has a minimum at that portfolio value.

I chose 77% as optimum, to prevent venture-bank jackpotting temptation while being fair to bankers.

The caution is that fiduciaries will be quite good at figuring out how to maximize for their own account, and any degree of improvement by damaging some of their portfolio investments is a perverse incentive to play to lose in order to make a little bit more off the . I suggest great caution in lowering either the clawback fraction below 75-77%, or setting the minimum term for an EDCS contract below 5 years. Should the clawback fraction be lowered, the minimum term needs to rise. If the minimum term drops then the clawback fraction needs to rise to defeat perverse incentive. Either that, or else the contract needs to allow the underwriter to implement a 100% penalty clawback if the underwriter finds fraud and has a claims process.

Figure 11 graphs in 3 dimensions what the effect of having no clawback would be on venture-bank earnings.

## 5.6 PREMIUMS PAID TO UNDERWRITER BY VENTURE-BANK

This summation is simplified by an assumption that all payouts occur in year 5, and all investment exits occur in year 10. So, we calculate the 5 and 10 year elements, and then multiply by the fraction of investments that apply to each.

*EDCSrate* is the EDCS yearly rate.

$Intrst = 0.02$

$EDCSrate = 0.05$

### 5.6.1 Underwriter side

*5 year term gross*

Total premiums if all was for a 5 year term.

5 Year term = 0.25

$$ProcYintercept1(P) \cdot DINrate \cdot 5 \qquad \textbf{(16)}$$

where: $P$ is the average return of the venture capital (VC) portfolio.

When P = 1.55, 5 year payments total = 0.1151163575

*10 year term gross*

Total premiums if all was paid for the full 10 year term.

10 Year term = 0.50

$$(1 - ProcYintercept1(P)) \cdot DINrate \cdot 10 \qquad \textbf{(17)}$$

where: $P$ is the average return of the venture capital (VC) portfolio.

When P = 1.55, 10 year payments total = 0.2697672850

Total payments = 5yr + 10yr (Underwriter side)
Payments = 0.1151163575 + 0.2697672850 = 0.3848836425

### 5.6.2 Venture-bank side

Venture-bank's version includes cost of money to carry.

$$ProcYintercept1(P) \cdot DINrate \cdot \int_0^5 (1 + Intrst)^{(5-x)} dx \; +$$
$$ProcYintercept1(P) \cdot DINrate \cdot \int_0^{10} (1 + Intrst)^{(10-x)} dx \qquad \textbf{(18)}$$

where: $P$ is the average return of the venture capital (VC) portfolio.

When P = 1.55, premiums & carry cost for 5 + 10 years is:
0.1210082154 + 0.2983317780 = 0.4193399934

## 5.7 EQUITY FRACTION OWED TO UNDERWRITERS

This is the fraction of the final equity of 1.55 that the underwriters have claim on. Since the underwriters have already taken the equity for all the companies with losses, the underwriters are only owed for the money making investments. So, the equity integral (eq. 19 & 20) lower bound is where the earnings function crosses 1 (e.g. where it crosses break even). Then the underwriter's fraction is multiplied by that fraction of total earnings.

I can see the possibility that underwriters may be tempted to lower their equity fraction in order to close sales to venture-banks for VC portfolio deals. Given that the overall performance of underwriter EDCS policies depends on the top end compensating for the low end, underwriters need to model the outcome.

A safe method of price cutting in a competitive underwriter market for venture-bank business would be to set a formula for equity fraction to slide based on value of a pool at exit. Such a formula should have a running accounting for the cost of payouts for the investment pool. This algorithm could be modeled such that it has good predictability. The only limitation is that when a payout or exit is triggered, the calculation must be executable in a transparent manner without giving rise to questions that might allow challenge to the status of EDCS instruments as derivatives. Consequently, making provisions for back end givebacks to venture-banks that meet certain performance targets is the proper way to structure such negotiaions.

### 5.7.1 Definite integral of equity fraction owed to underwriter

*EDCSEquityFraction* = 0.5

$$DINEquityFraction \cdot \int_{ProcYintercept\,1(P)}^{1} P - 1.55 + .2655 \cdot e^{(2.88 \cdot h)}\, dh \qquad \textbf{(19)}$$

where: *P* is the average return of the venture capital (VC) portfolio; *e* is number *e*.
When P = 1.55, underwriter equity = 0.6475155435

### 5.7.2 Definite integral of equity fraction remainder for venture bank

$$(1 - DINEquityFraction) \cdot \int_{ProcYintercept\,1(P)}^{1} P - 1.55 + .2655 \cdot e^{(2.88 \cdot h)}\, dh \qquad \textbf{(20)}$$

where: *P* is the average return of the venture capital (VC) portfolio; *e* is number *e*.
When P = 1.55, Venture-Bank equity = 0.6475155435

### 5.7.3 Results for a VC portfolio of 1.55 (Kauffman's)

**Venture bankers per multiple of capital**

```
   0.6475155435    Venture-Bank equity
 + 0.2054307077    EDCS  payouts
 - 0.4193399934    Total EDCS premiums plus carrying cost
 - 0.1581816449    Clawback
==============
   0.275424613     Gain
```

$$Venture\,bank\,equity - (Total\,DIN\,premiums + carrying\,cost) - Clawback = VC\,earnings$$

$$VC\,earnings \cdot MOC = Venture\,bank\,ROI \qquad \textbf{(21, detail of 1)}$$

**Underwriters per subset portfolio**

```
   0.6475155435    Underwriter's equity
 - 0.2279261069    EDCS  payouts total carrying cost
 + 0.3848836425    Total EDCS payments
 + 0.1581816449    Clawback
==============
   0.962654724     Gain
```

**Underwriter's ROI**

$$\frac{Underwriter's\,equity + Total\,DIN\,premiums + Clawback}{(Insurance\,payouts + carrying\,cost)} = ROI \qquad \textbf{(22, detail of 2)}$$

$$\frac{0.6475155435 + 0.3848836425 + 0.1581816449}{0.2279261069} = 5.2235$$

### 5.7.4 Results for a VC portfolio of zero (-3 on these graphs)

**Venture-bank**

We repeat the above calculation with a portfolio adjustment of -3.
Total earnings for venture-bankers per multiple of capital

| | |
|---|---|
| 0.0 | Venture-Bank equity |
| + 0.9987995023 | EDCS payouts |
| - 0.2627954404 | Total EDCS payments and carrying cost |
| - 0.7690756168 | Clawback |
| ============== | |
| **(0.095811246 )** | Loss |

**Underwriters**

| | |
|---|---|
| 0.0 | Underwriter's equity |
| - 1.108171629 | EDCS payouts total carrying cost |
| + 0.25 | Total EDCS payment income |
| + 0.7690756168 | Clawback |
| ============== | |
| **(0.089096012)** | Loss |

# 6 VENTURE-BANK SYSTEM BEHAVIOR

## 6.1 WHEN ZERO ISN'T ZERO: *P* SCALE < 1 IS AN ADJUSTMENT FACTOR

The graph of the venture-bank results when the average VC portfolio return = 0 needs discussion. It would seem that a miracle occurs here, since you cannot squeeze something from nothing. And in the graphs below, you will see VC portfolio values down to -3. So why is this?

When the portfolio adjustment is zero, this would be a true zero portfolio if there was no floor on losses and investments could go negative without limit. However, in this model, investments can go to zero, but not past zero. (See figure 3.) Because of this, when the portfolio adjustment is at 0, the actual VC portfolio return is 0.5 (50%), or a 50% loss of invested capital. (See figure 6 for detail of behavior below the 0 portfolio adjustment factor.) To get to an actual return for a portfolio that equals zero, it is necessary to have a -3.17 portfolio adjustment[3]. However, here I have used -3 as the lowest value because it is functionally the same as a VC portfolio that goes to true zero out to 3 decimal places, and any venture-bank that actually manages this feat has other serious problems.

From a value of 1 up, the adjustment scale x-axis is the same as the real scale.

I think we are mostly interested in the region from 0 (0.50) to 1.875, because this shows behavior for those large portfolios that should be most common. A portfolio factor of 0 is a 50% loss portfolio, and yet both parties can still make money at this level. I assume a VC portfolio that averages losses of 50% is quite poor. In the Kauffman dataset, 0.5 is in the bottom octile. Similarly, 1.875 is about as good as any large VC portfolio is likely to get with current VC selection approaches. One might then ask, why I show up to a +3 VC portfolio return. The reason is that Mulcahy noted that VC firms have a correlation between past performance and future results that is not seen in the stock market. So, it is plausible to think it may be possible to achieve large portfolio

3 While it is possible to create a function that would take a value of 0 and turn it into a -3 internally, so that the graph scale shows the true VC portfolio return on the horizontal axis, this creates extra complexity internally, and greatly increases the execution time. At this stage, I decided to avoid that. I

VC results greater than 1.875 in the future with enough study.

Remember that the purpose of a venture-bank is to create money for the venture-bank to loan to themselves. This, and the fact that the investment equity remains the asset of the bank is key to how venture-banking returns work.

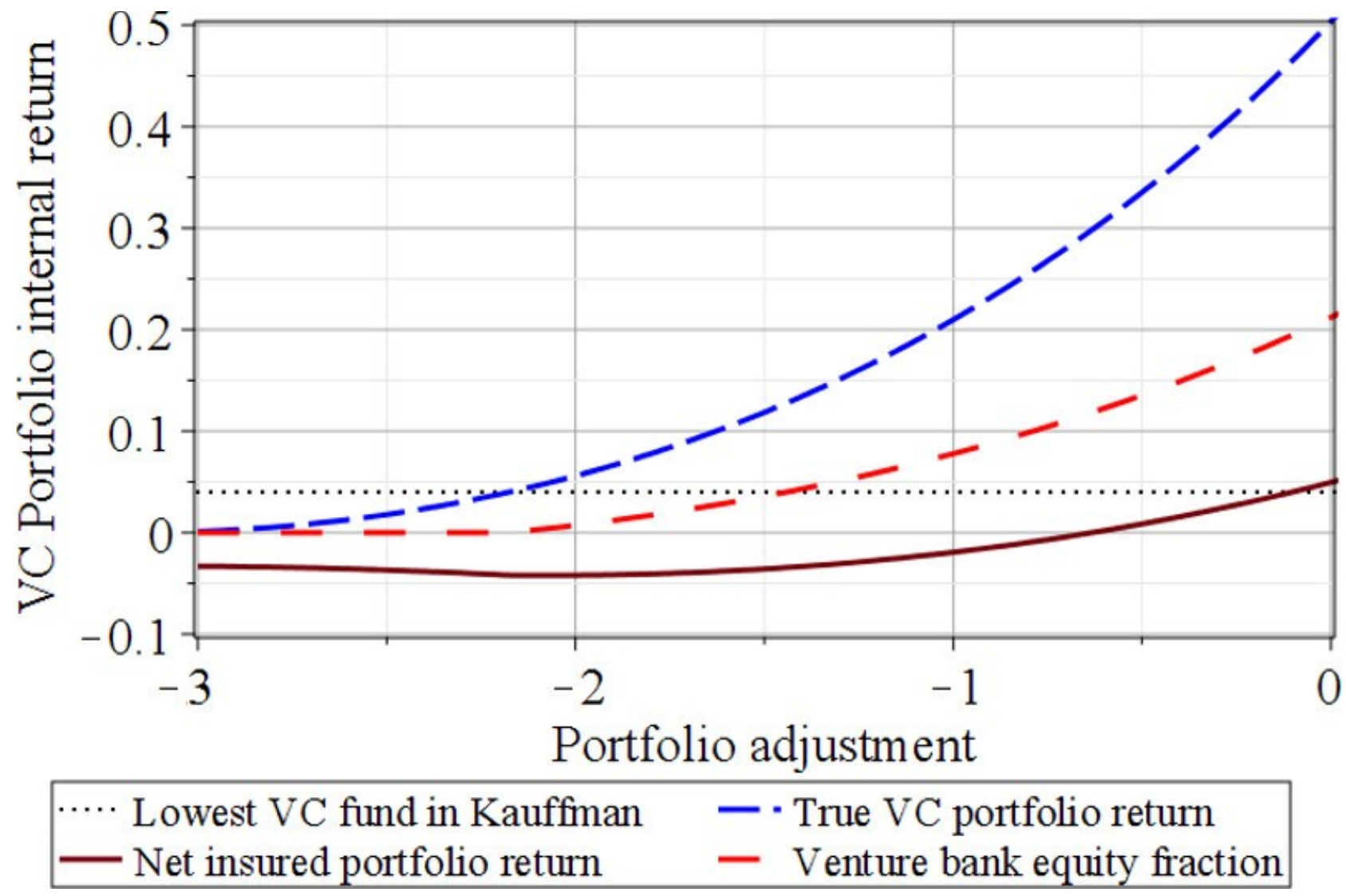

Figure 6: Detail for individual deal returns showing region of low or negative net. The blue dashed line is the results of the investment total return equation, what the VC portfolio return is without the venture-banking mechanisms. Zero on the portfolio adjustment scale is a portfolio with 50% losses. Portfolio losses go to zero at an adjustment of -3. The lowest Kauffman fund VC fund returned 0.04 shown by dotted line. The blue line crosses 0.04 at approximately -2.18. The red dashed line shows the venture-bank's equity share return. The black solid line shows the net return with EDCS premium payments, clawbacks and payouts.

I show the system of equations behavior down to -3. Be aware that -3 is not quite a 100% loss VC portfolio, and 0 is a 50% loss VC portfolio. However, a breakeven portfolio of 1.0 is correct, as is everything larger. The function is non-linear, and converges to be correct near 1. I chose not to correct this non-linear element of the model at this time to prevent complicating its internals.

## 6.2 VENTURE-BANK GRAPH RESULTS

Venture-Bank ROI =

$$(Venture\ Bank\ equity + Insurance\ payouts - Premium\ payments\ cost - Clawback) \cdot MOC$$

**(23, restatement of 21)**

Figures 7 through 12 are views of the same results, rotated around the X, Y and Z axes.

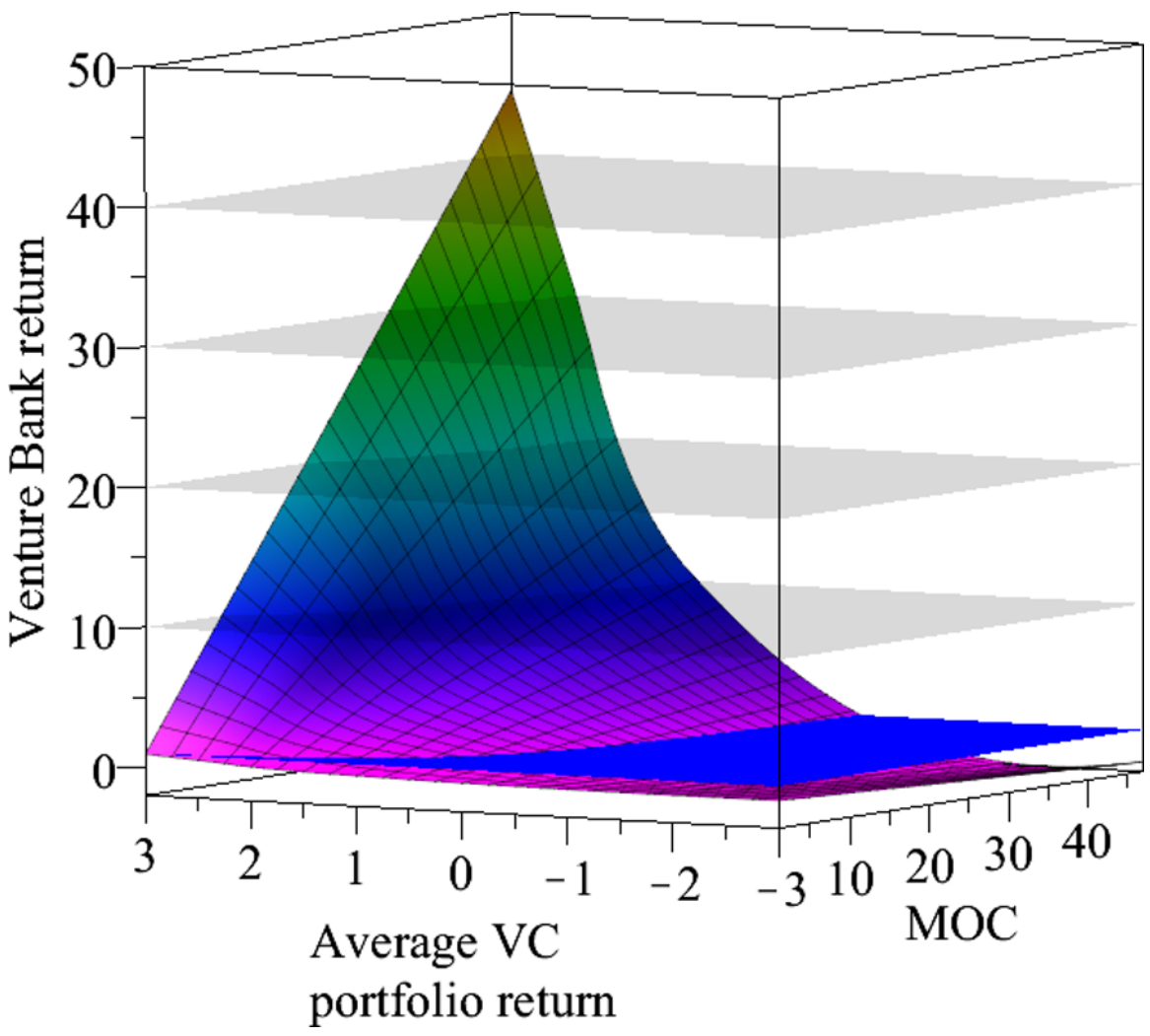

Figure 7: Venture-bank returns. Blue plane is at 1, which is breakeven for the bank

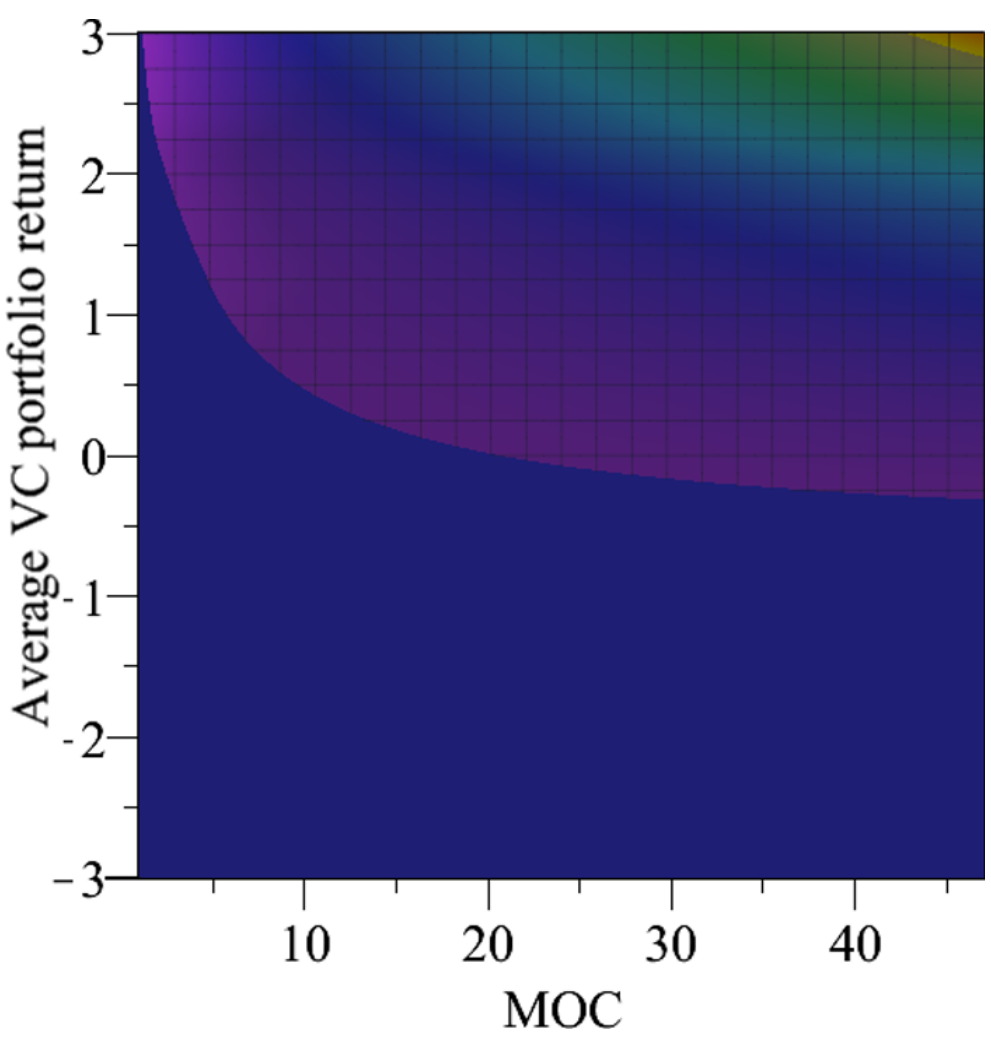

Figure 8: Top view. Blue plane is breakeven for the bank.

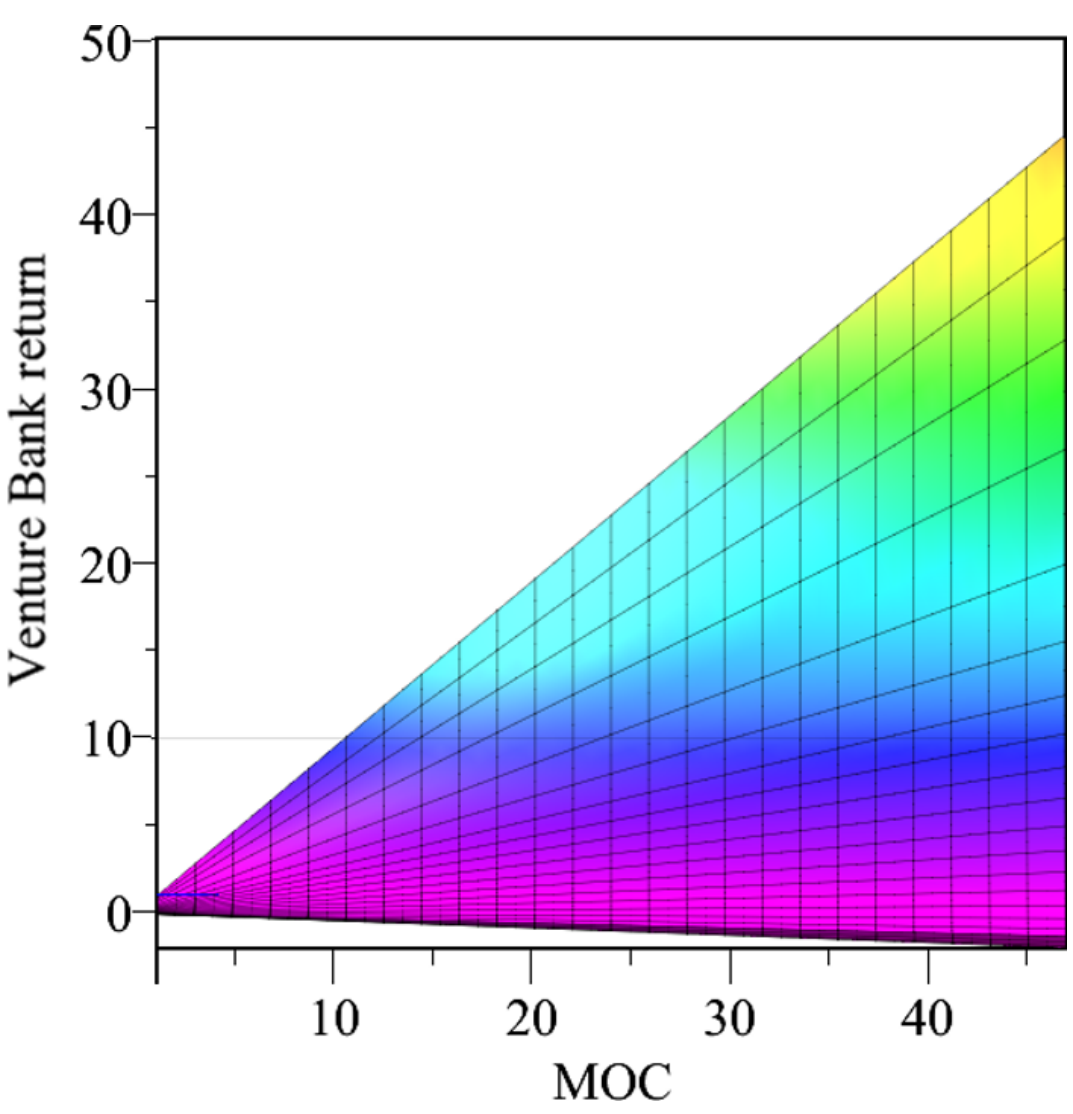

Figure 9: Side view along VC portfolio axis

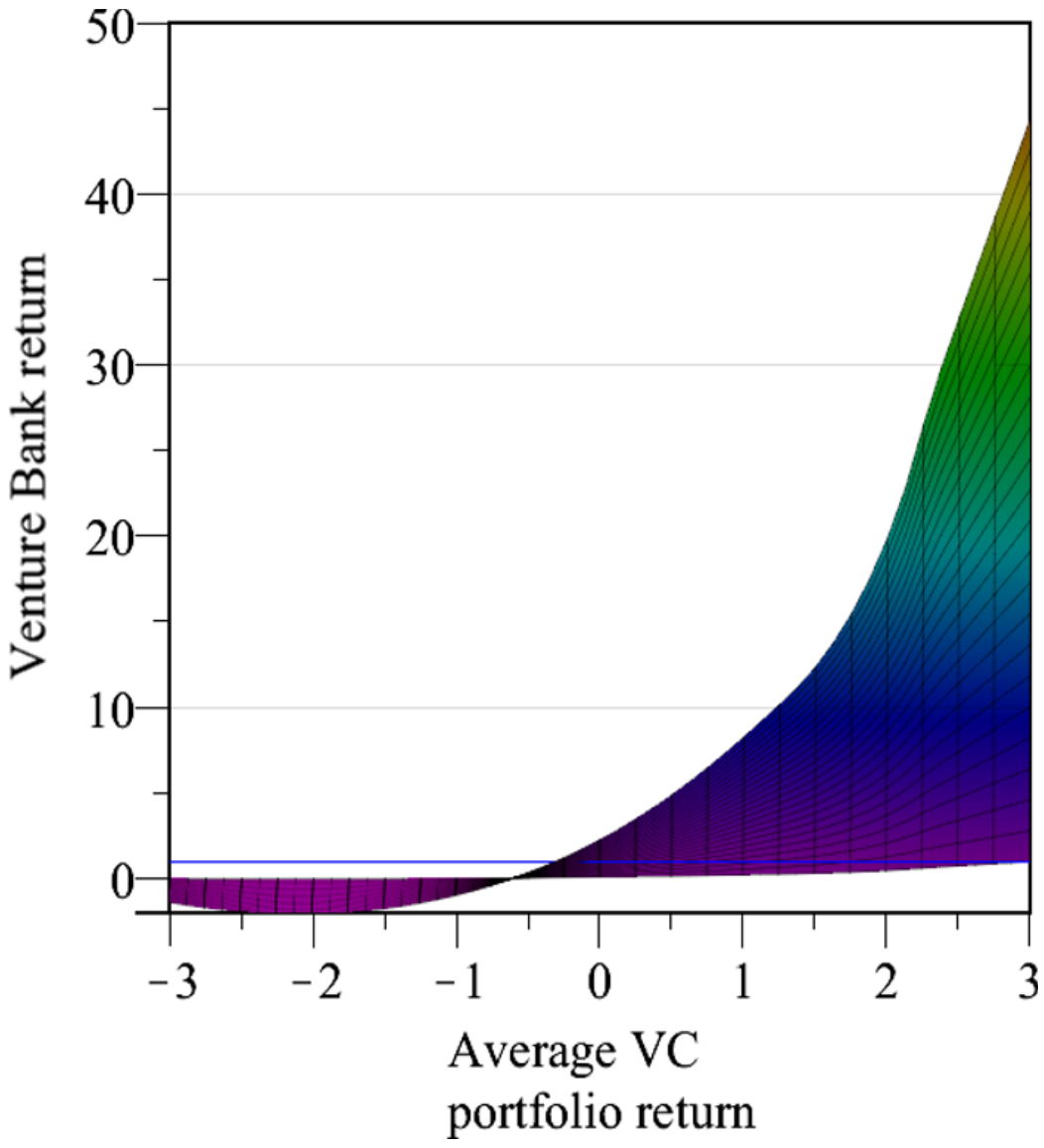

Figure 10: Side view along MOC axis. Blue line at 1 is breakevn.

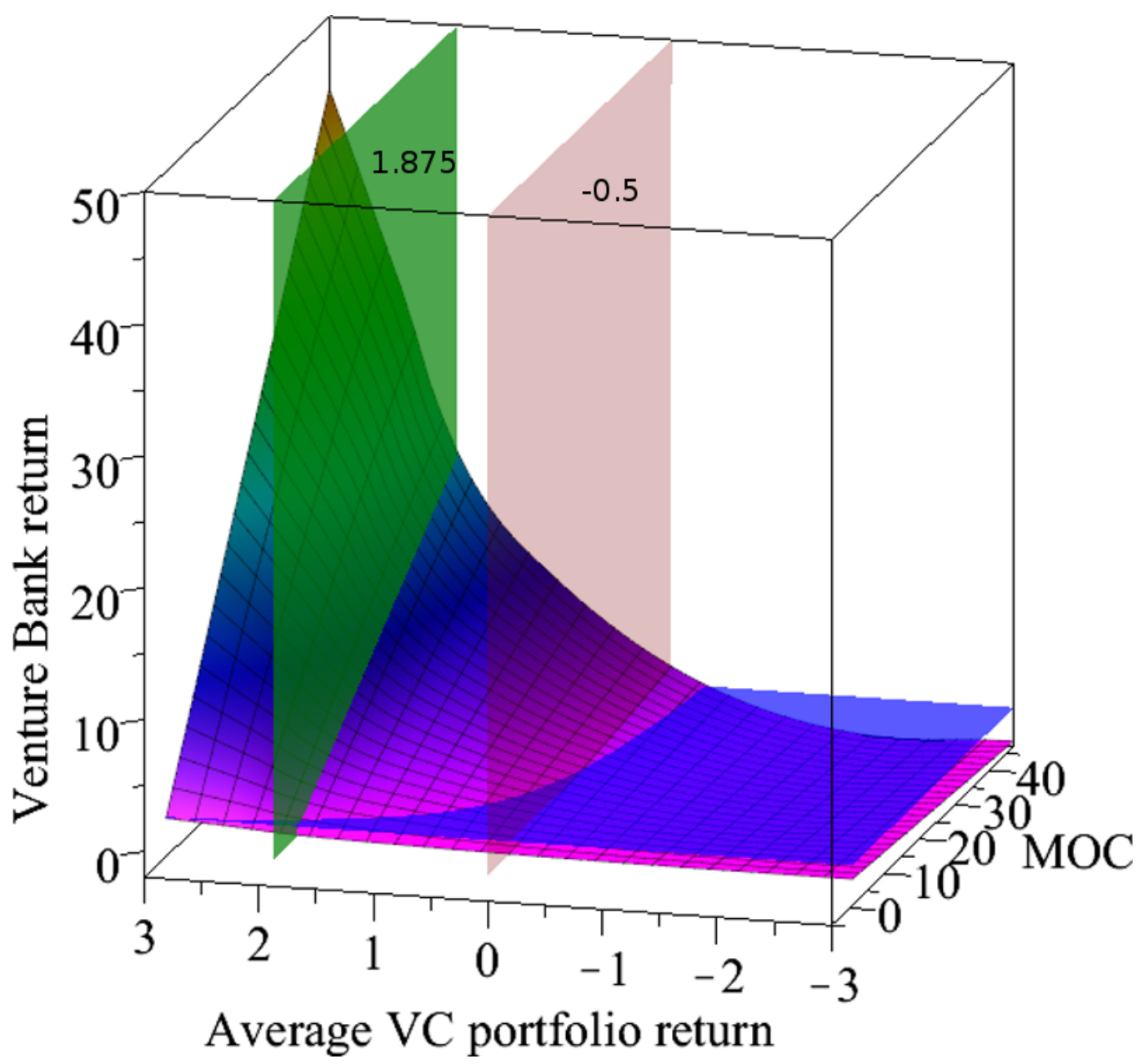

Figure 11: Venture-bank returns. Red cut plane at zero marks returns that correspond to a 50% loss VC portfolio. Green cut plane correctly shows 1.875 VC portfolio return. Blue horizontal cut plane at 1 is breakeven.

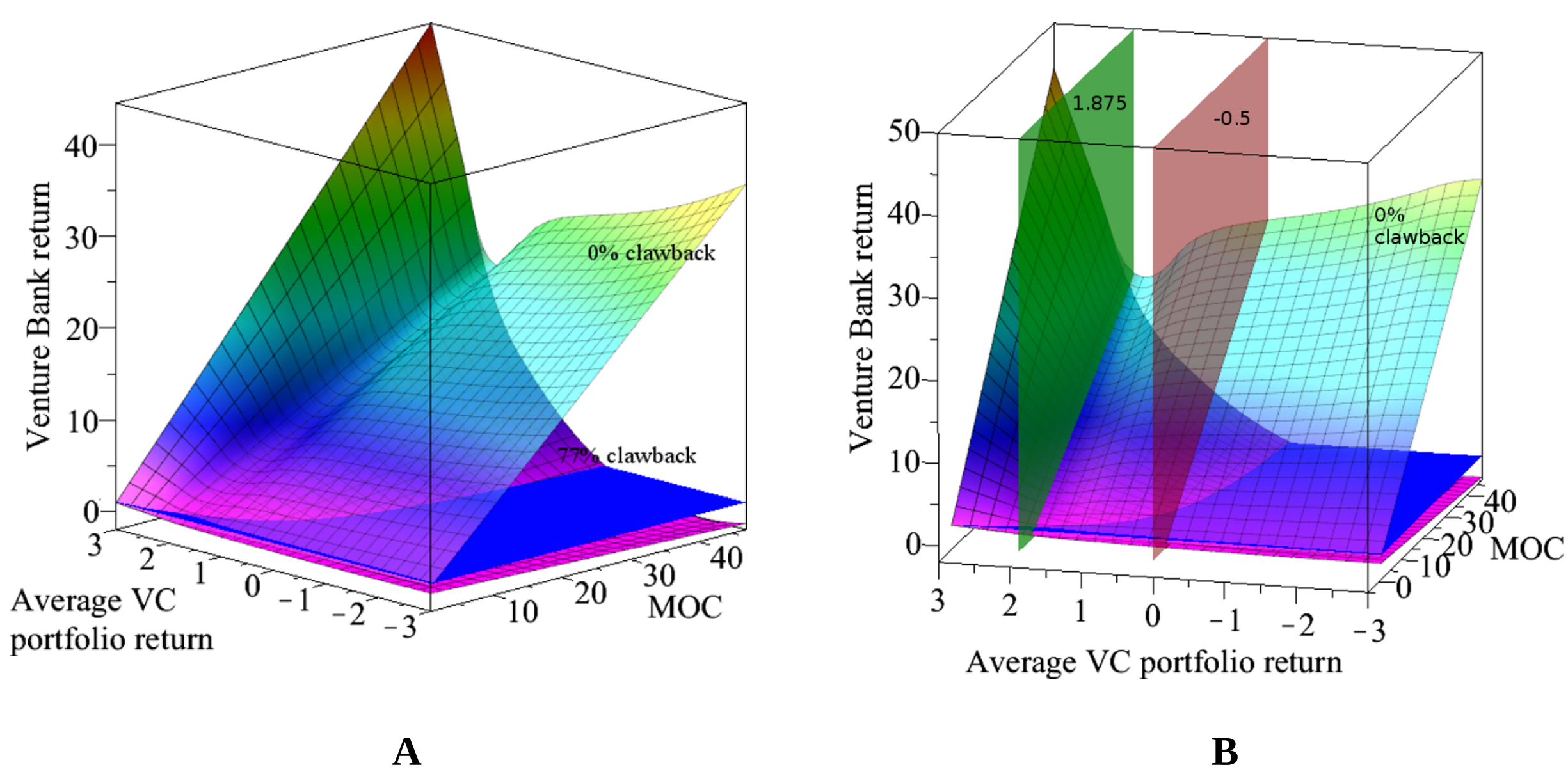

**A** **B**

Figure 12 A & B: Behavior with 0% clawback versus 77% clawback. This figure shows why it is necessary to have clawbacks. Otherwise, there is a valley in the region where most of the VC portfolios are. Most venture capital portfolios should be in the region between 0 and 1.875 bounded by the red and green cut planes. The red cut plane at zero corresponds to VC portfolio returns of -0.5 (50% loss). The green cut plane at 1.875 correctly identifies a VC portfolio return of 1.875 (187.5%).

## 7 UNDERWRITER SYSTEM BEHAVIOR

For the underwriter, MOC is not meaningful relative to ROI. Instead, MOC is related to the total volume of business that can be done with the bank. Let us step through the basic business of the underwriters.

Earnings are:

Total premiums to underwriter + EDCS Equity share + Clawback – (Payout + Carry cost).

There are four ways of figuring the return on investment.

1. Investment is the cost of payout plus the cost of money to pay for it carried over 5 years.
   Payout + Carry cost.
   Earnings are: EDCS premiums + Equity + Clawback
2. Investment is Payout + Carry cost - Discounted clawback bond sales.
   In this scenario, clawbacks, or some fraction of them, are sold, to defray costs for Payout + Carry.
   The discount rate of clawbacks will be set at 1.5 x USA 12 month LIBOR.
3. Investment is Payout + Carry cost - Sales of up to 70% of discounted equity futures.
   EDCS Equity future shares for this model are discounted at 1.5 x (USA 12 month LIBOR) from the average portfolio valuation at exit.
   Note: Management of EDCS equity shares will be the responsibility of the underwriter at closeout, or an alternative exercise competent entity that takes over 30% or greater majority position (Hanley 2017b). (Underwriter cannot drop responsibility for being exercise competent entity without another shareholder with 30% or more agreeing to become one.) With an exercise competent entity taking over equity futures, the underwriter can sell 100% of EDCS equity futures, but that is not modeled here.
4. Combination of sales of clawbacks at discount and sales of equity futures.
   Here, investment is:
   Payout + Carry cost - Discounted clawback - Discounted equity futures sales.

Figure 13 shows behavior of components of ROI for underwriters

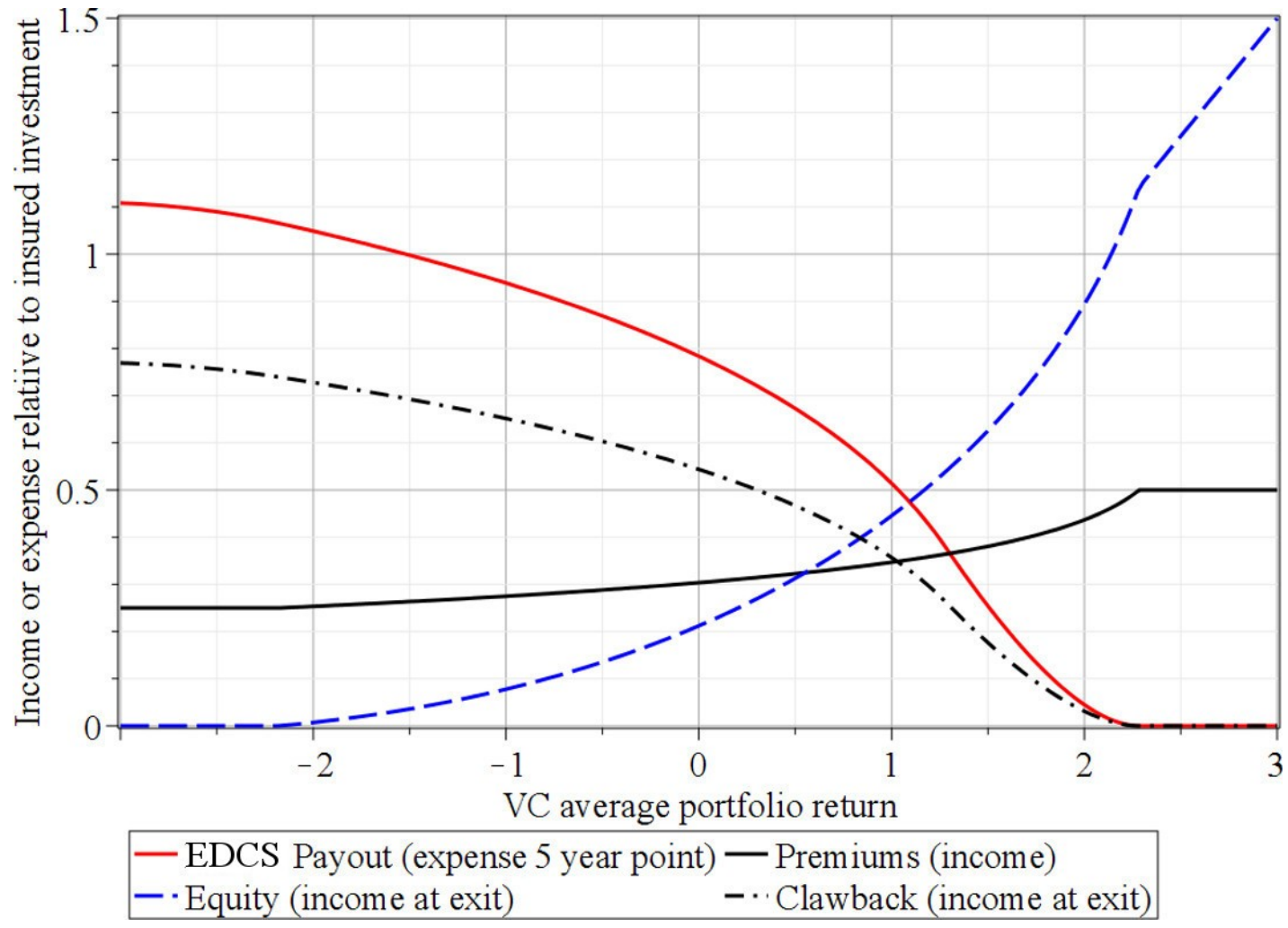

Figure 13: Cost and income components of underwriter returns. Solid descending red line is underwriter payouts which decline as portfolio profitability rises. Dash-dot descending black line is payout clawback returned to underwriter which declines here as 77% of the payout. Dashed ascending blue line is underwriter's equity which here is 50% of total equity generated. Solid black ascending line is EDCS premium income, which has a flat minimum where all investments return less than 1, and so within our assumptions, all investments only pay premiums for 5 years. EDCS premium income rises until all investments return 1 or greater, and so all pay premiums for a full 10 years.

## 7.1 1. SIMPLE UNDERWRITER ROI

Underwriter ROI =

$$\frac{InsNetDINPremiumTotal + InsEquityShare + InsNetClawbackLien}{InsLossPayoutTotal} \quad \textbf{(24)}$$

Figure 14 shows the ROI if the underwriter does not sell equity futures or clawbacks at a discount.

Underwriters losses begin at -1.4525, which corresponds to a VC portfolio return of 0.125975915 (~12.6%, or a portfolio loss of 87.4%).

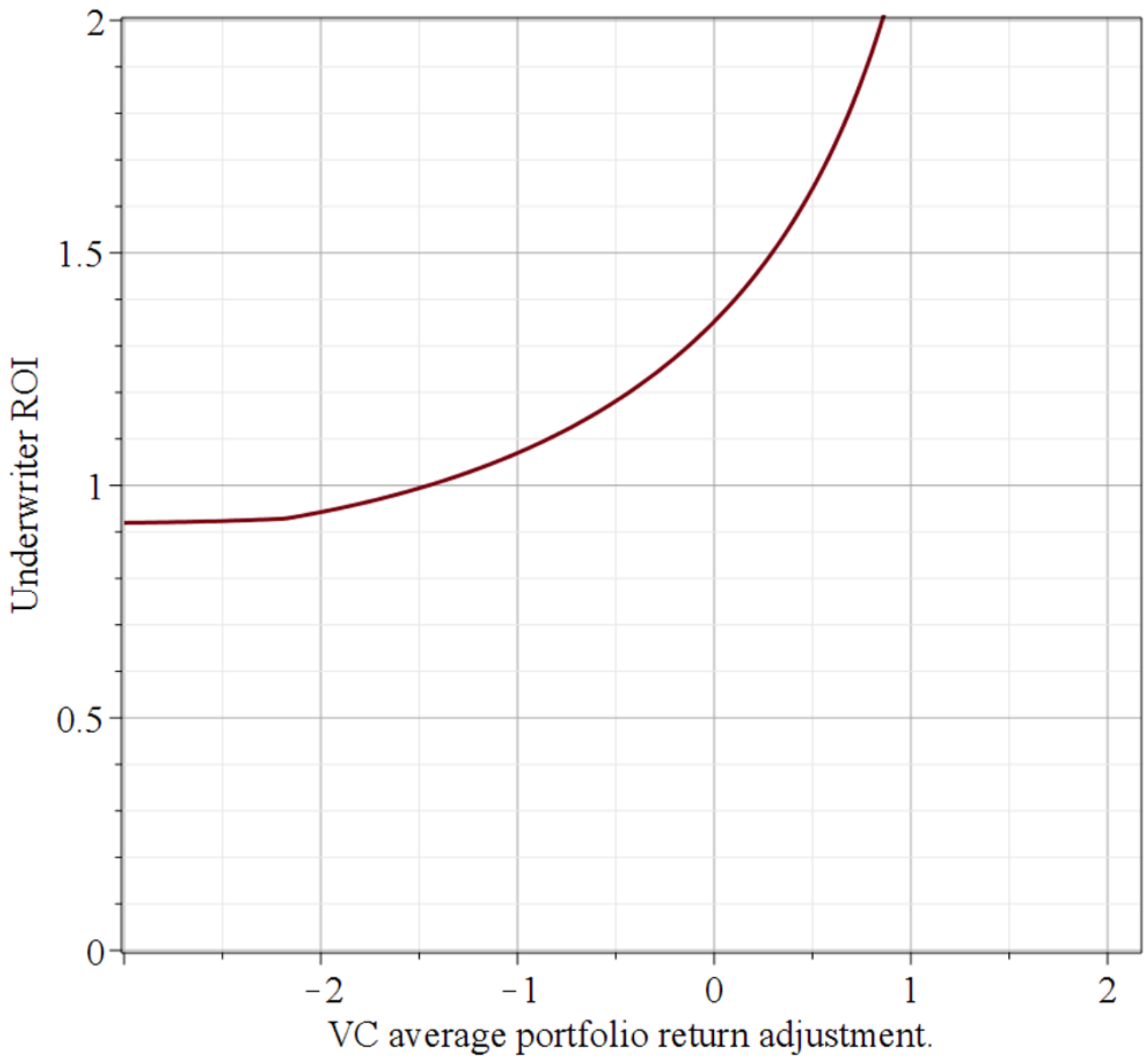

Figure 14: Detail view of simple underwriter earnings without sales of clawbacks or equity futures. ROI becomes meaningless where VC portfolio scale is above 2. Breakeven is at -1.425

### 7.2 2. UNDERWRITER ROI WHEN SELLING OFF CLAWBACK BONDS

Underwriter ROI =

$$\frac{InsNetDINPremiumTotal + InsEquityShare + RemainderInsNetClawbackLien}{InsLossPayoutTotal - DiscountedInsClawbackLienSales} \tag{25}$$

The idea here is that shares in clawback liens would be sold off as fixed maturity bonds.

Figure 15 shows the changes in ROI as the fraction of discounted clawbacks sold changes for a selected set of portfolio adjustments. Note that with clawback bond sales, the portfolio adjustment break-even point of -1.425 in the previous simple accounting is only stable at approximately -0.25.

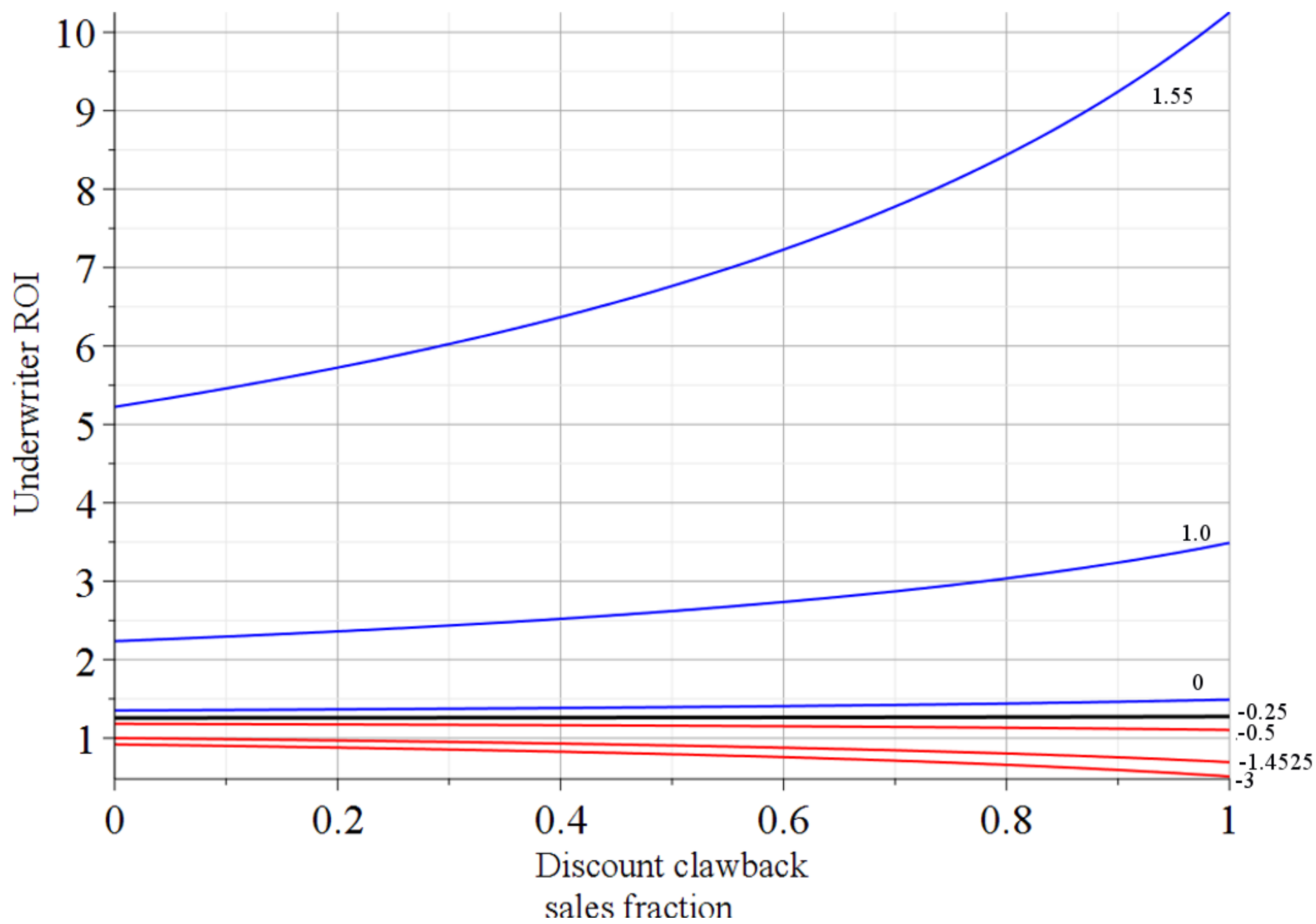

Figure 15: VC portfolio adjustment numbers are shown next to the lines on the right. Note that here, a portfolio adjustment of -0.25 (black line) is flat. Everything below this, ROI trends down the larger the clawbacks fraction sold. Everything above it, the ROI rises the more are sold. Compare with figure 23.

Figure 16 shows how ROI varies with both clawback bond sales and adjustments of the portfolio returns for individual deals. The blue plane is at 1, which is break even. The zero clawback sales graph in gray is shown together with the color shading sales graph. When the color shaded graph ends while ascending, ROI becomes meaningless due to zero net cost of payouts.

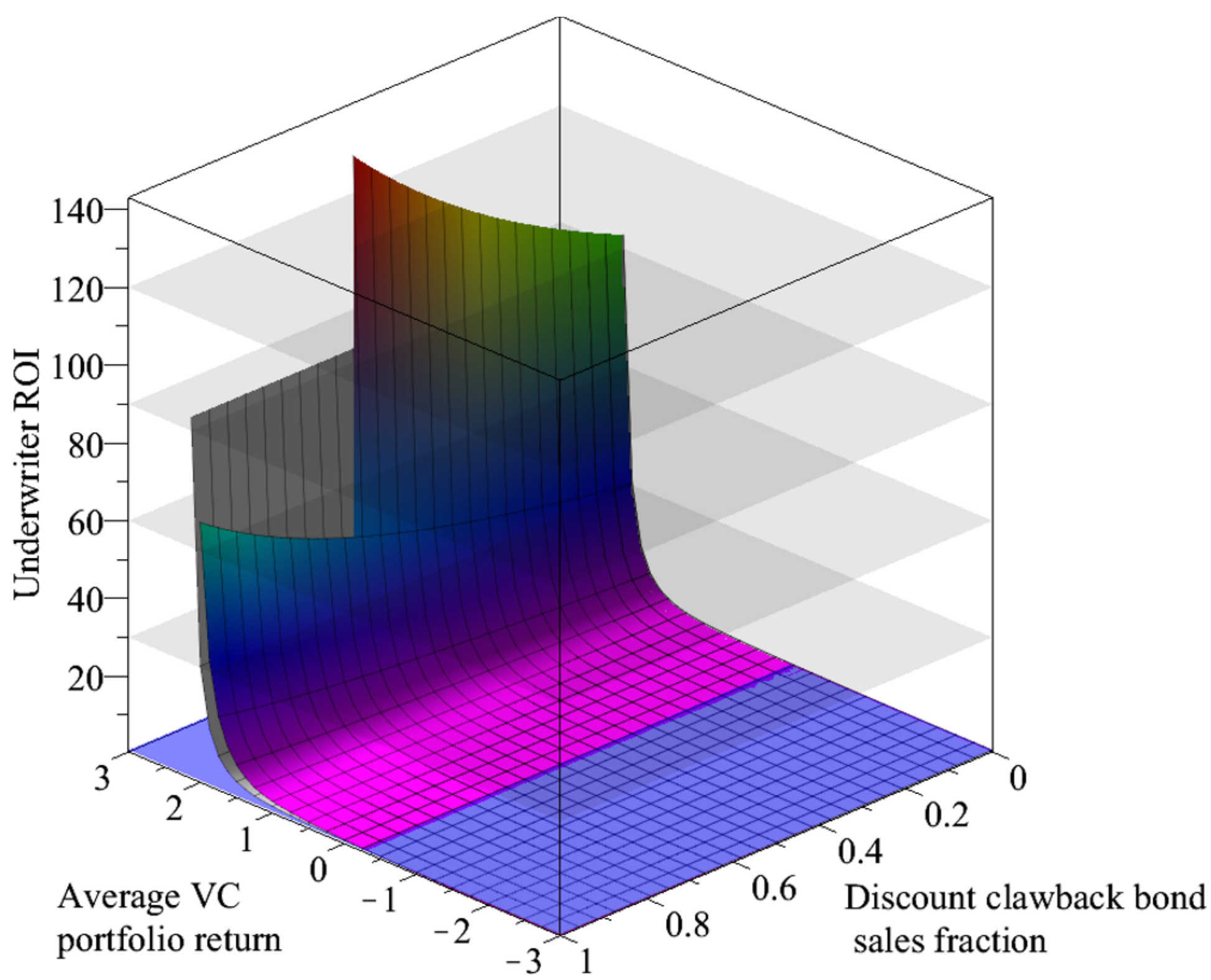

Figure 16: Varying sales of discounted clawback bonds. With sales, ROI rises a little faster above the -0.25 transition. But below -0.25, it drops below break-even, and has worse results than without. Where the graph ends going vertical, ROI is infinite because of dividing by zero cost.

## 7.3 3. UNDERWRITER ROI WITH SALES OF EQUITY FUTURES

Underwriter ROI =

$$\frac{InsNetDINPremiumTotal + InsNetClawbackLien + RemainderDiscountedInsEquitySales}{InsLossPayoutTotal - DiscountedInsEquitySales} \quad \textbf{(26)}$$

Figures 17 and 18 show how ROI varies with equity futures sales while the VC portfolio adjustment goes from -3 to 3. In the region of higher VC portfolio return in figure 16 beyond the spikes, there is no cost, only earnings.

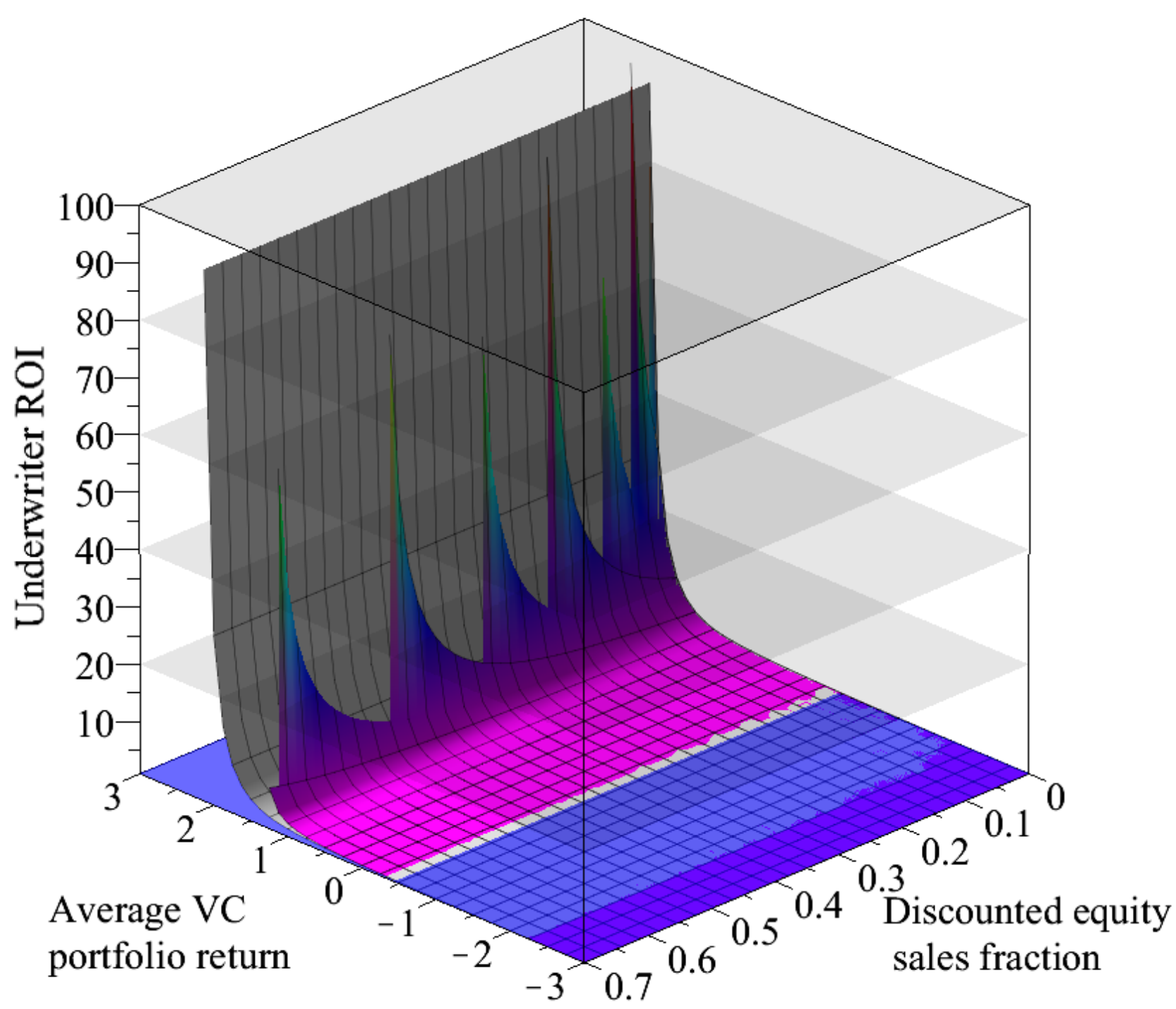

Figure 17: Comparison of sales of equity futures with no sales (in gray). The significance of this graph is, again, that ROI rises faster above the critical transition, and below it, drops more quickly.

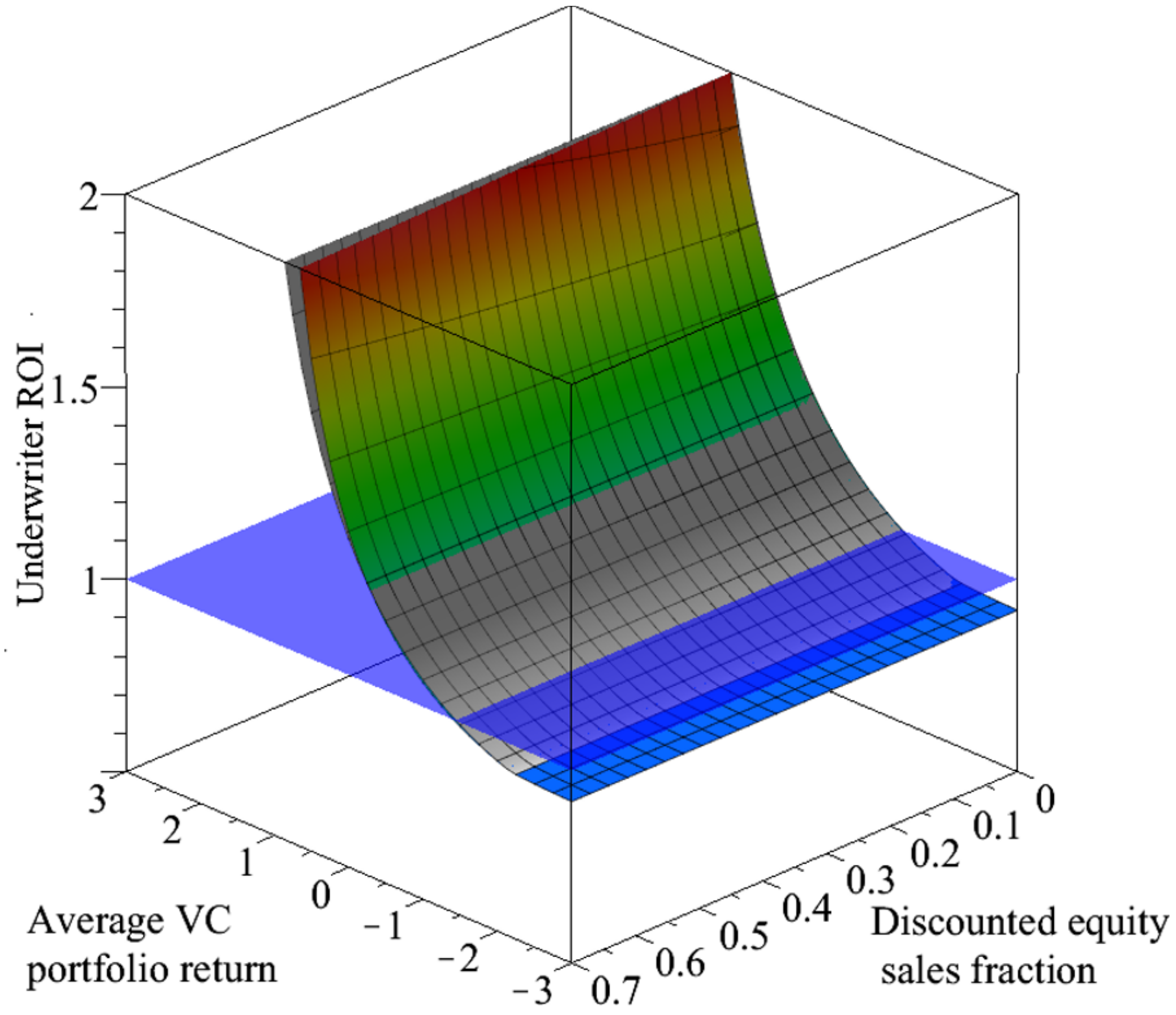

Figure 18: Detail of discounted equity futures sales. Here we see the transition region more clearly as the color shaded sales graph dips dips below the gray non-sales graph.

## 7.4 4. UNDERWRITER ROI WITH SALES OF BOTH CLAWBACK BONDS AND EQUITY FUTURES

Underwriter ROI =

$$\frac{InsNetDINPremiumTotal + RemainderInsNetClawbackLien + RemainderDiscountedInsEquitySales}{InsLossPayoutTotal - DiscountedInsClawbackLienSales - DiscountedInsEquitySales} \quad \textbf{(27)}$$

Figures 19-22 vary the VC portfolio adjustment from -3 ( zero return) to +2. Above 2.14 there is no graph, which means there is no net cost to the business.

The region past where the graph spikes end is all earnings and ROI is meaningless. Remember that a VC portfolio adjustment of 1 is 1, and everything above that is likewise correct. But below 1, the portfolio adjustment becomes inaccurate. A portfolio adjustment of 0 is a VC portfolio return of -0.5, or a 50% loss. A portfolio adjustment of -3, is a VC portfolio of zero.

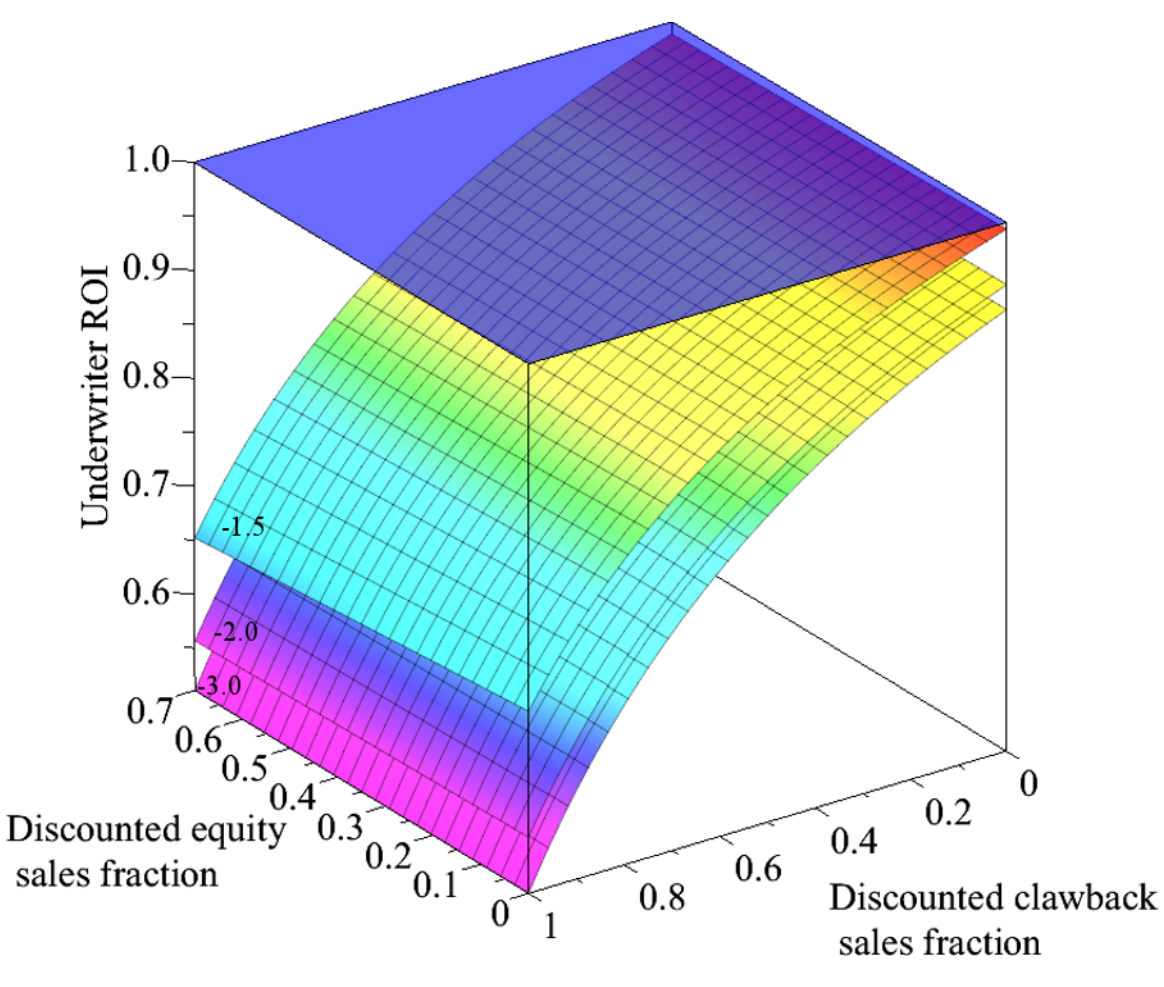

Figure 19: VC portfolio adjustment = -3 to -1.5. Blue cut plan at breakeven.

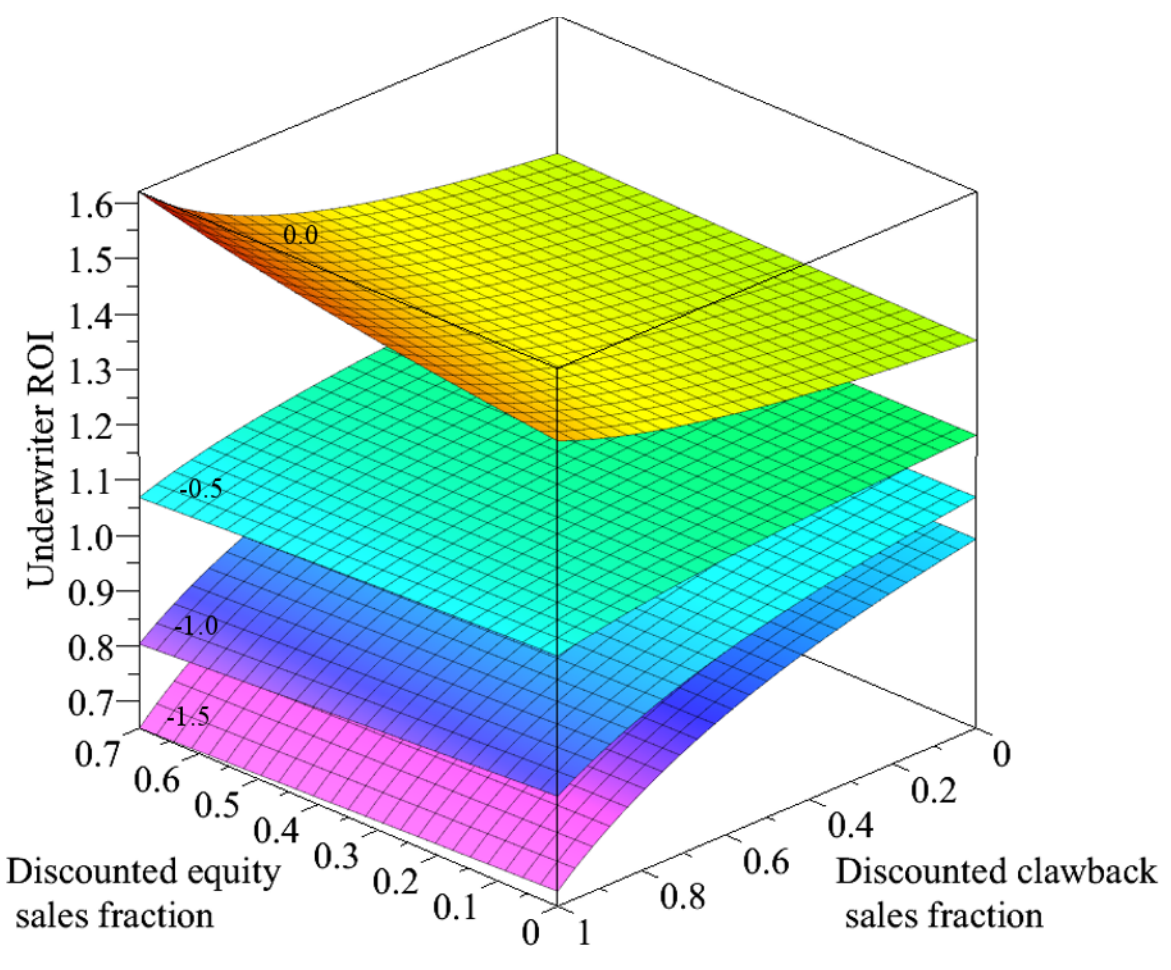

Figure 20: VC portfolio adjustment = -1.5 to 0.0.

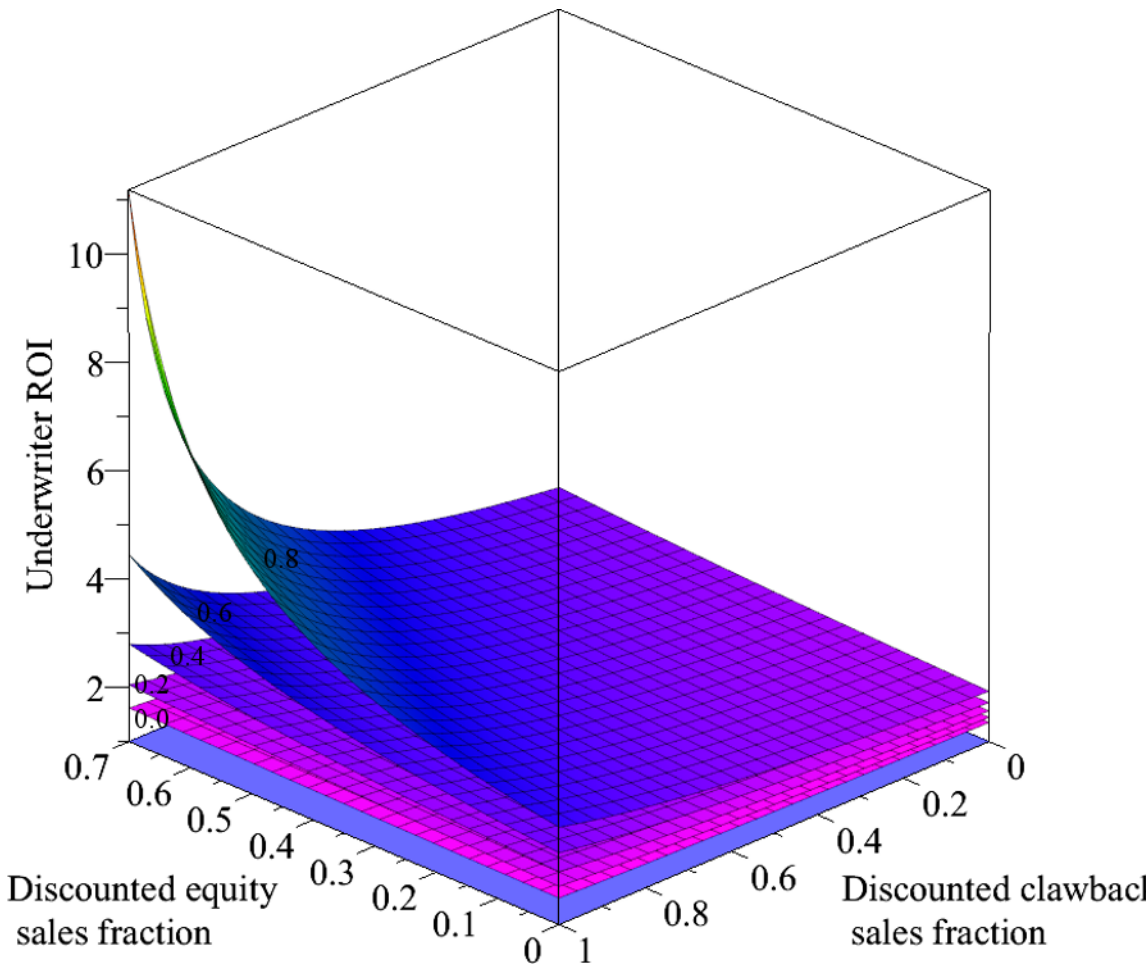

Figure 21: VC portfolio adjustment = 0.0 to 0.8. Blue cut plane at breakeven.

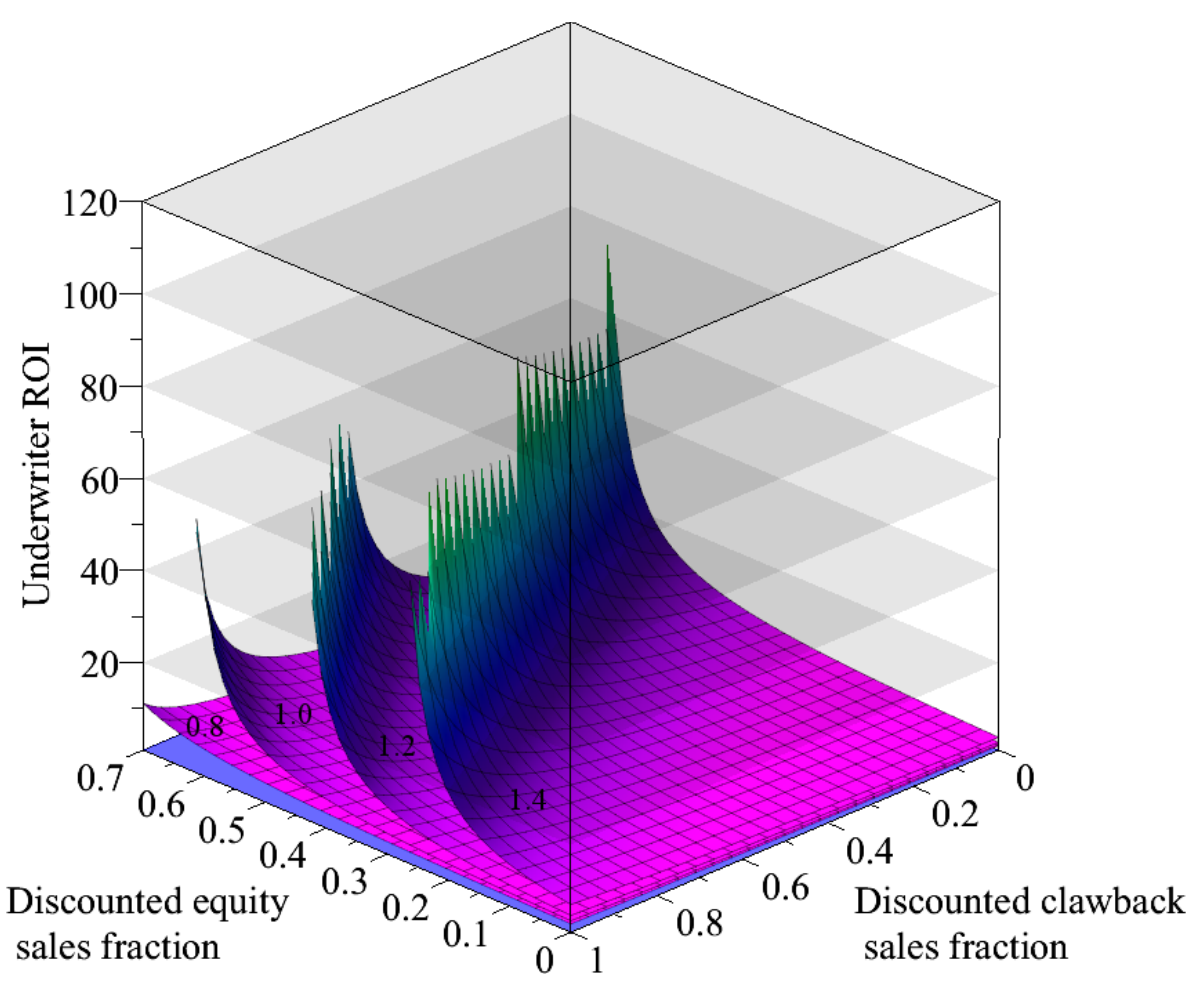

Figure 22: VC portfolio adjustment = 0.8 to 1.4.

## 7.5 COMBINED GRAPHS FOR UNDERWRITER SALES OF BOTH CLAWBACK BONDS AND EQUITY FUTURES

Figures 23 and 24 show how ROI varies based on cash flow as affected by sales of clawback bonds and discounted equity futures at selected VC portfolio adjustments together in one graph.

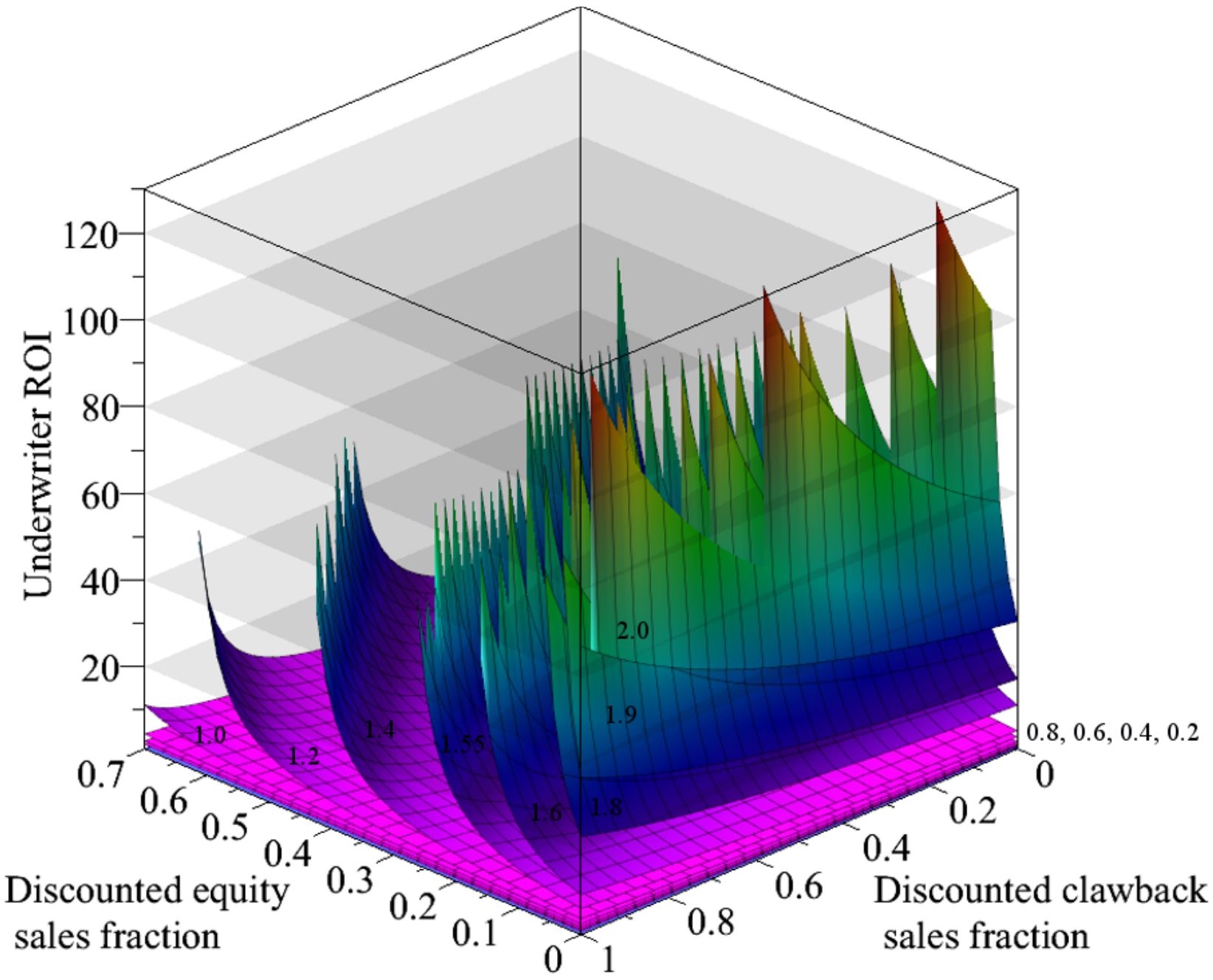

Figure 23: VC portfolio adjustment = 0.2 to 2.0

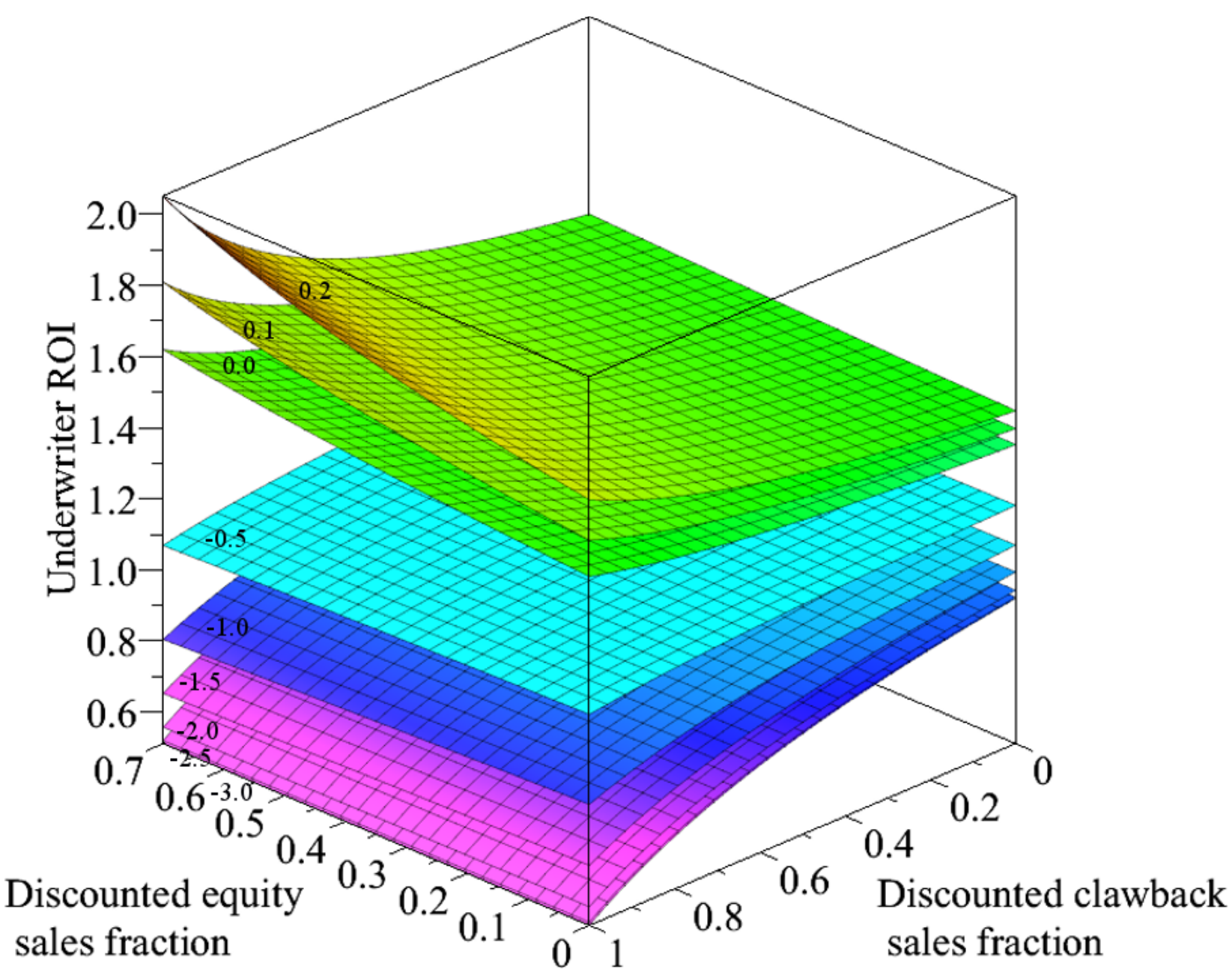

Figure 24: VC portfolio adjustment = -3 to 0.2

## 8 CONCLUDING REMARKS

In a reasonably normal operating environment, the optimum strategy for the underwriters will be to sell EDCS and clawback futures off early. This will generate revenue, and insulate the underwriter from markets over time. There is plenty of time to do this, so underwriters should be in a good position to establish a market for new issues. Because of this, I would expect that equity futures should usually fetch a significantly better price than the perfect fore-knowledge calculation of this model.

A further benefit of sales of equity futures is that these can be used to establish a surrogate pricing for the underlying investments. If this can be established, it may be usable for creating alternative exits, such as leveraged buy outs for those small number of investments that are not ready to retire at 10 years. Such transactions would have to be done outside of the venture-banking system, but with equity futures pricing available, this may help.

If all the clawbacks are sold off then the underwriter loses transparency into the venture-bank's records relative to that investment. If there is any reason to want to maintain transparency, then that should be avoided. Like the equity futures, this should significantly improve ROI, and provide an earlier conversion to an effectively zero cost business.

I expect that payoff occurrences will exhibit clustering, and should skew toward early payoffs for three reasons.

- First, Kauffman's data shows that funds exhibiting losses are concentrated in a rough quartile, and that past performance of a fund's managers was a good predictor of future performance.
- Second, when an EDCS is held, VC managers will have less incentive to hold investments that they think won't get to break even because it is costing them the premium to maintain the investment. This will be balanced to some extent by the requirement to keep paying on the policy for a minimum term.
- Third, my assumption that VCs only take the 20% carry on their positive gain portfolios is a conservative assumption. The true venture capital performance internally may be significantly higher.

There will be performance impact for individual investment returns on venture-bank portfolios due to paying for EDCS contracts on investments. Against the incentive to cut losses earlier due to maintenance of EDCS contracts, the venture-bank will want to maximize the value of that investment in order to minimize losses because of the clawback lien credits. I think that for the venture-banks, this will probably improve performance relative to current VC firms.

It will be necessary for an underwriter to set minimum terms, I suggest 5 years, to prevent dumping early. Otherwise, underwriters could get hit with some of their clients paying 1 year of EDCS premiums and dumping investments early in order to minimize their EDCS costs while maximizing payoffs on dumped investments. This scenario would damage underwriter's returns, and has potential to undermine the stability of the venture-bank system. Against this, if underwriters maintain a back end claims process during some portion of the clawback period, such activity can be treated as a violation of fiduciary trust, and result in a higher clawback up to 100%, and possibly could include penalties above and beyond the clawback amount.

# 9 APPENDIX

## 9.1 KAUFFMAN DATASET

Octiles marked with shading.

*Dataset 1 – Original Kauffman dataset*

Vector([0.04, .10, .10, .15, .15, .20, .30, .30, .30, .30, .40, .50, .60, .60, .60, .65, .65, .65, .65, .65, .65, .65, .70, .70, .70, .70, .75, .75, .75, .75, .75, .75, .80, .80, .85, .90, .90, .90, .90, .90, .90, .90, .90, .90, .90, .90, .90, .99, .99, .99, 1.05, 1.10, 1.10, 1.10, 1.10, 1.10, 1.20, 1.20, 1.24, 1.25, 1.25, 1.30, 1.30, 1.30, 1.30, 1.35, 1.35, 1.35, 1.35, 1.35, 1.35, 1.40, 1.50, 1.50, 1.50, 1.60, 1.60, 1.70, 1.70, 1.70, 1.70, 1.70, 1.80, 2.10, 2.10, 2.20, 2.20, 2.20, 2.20, 2.30, 2.30, 2.30, 2.60, 3.00, 3.20, 3.20, 3.20, 3.80, 6.00, 8.00], datatype = float)  Mean average = 1.31

| | **1** | **2** | **3** | **4** | **5** | **6** | **7** | **8** |
|---|---|---|---|---|---|---|---|---|
| **Octile mean** | 0.24 | 0.65 | 0.79 | 0.92 | 1.17 | 1.38 | 1.84 | 3.41 |
| | **1** | **2** | **3** | **4** | | | | |
| **Quartile mean** | 0.452 | 0.857 | 1.276 | 2.656 | | | | |

## 9.2 KAUFFMAN REVISED DATASET

Values are identical until they are greater than or equal to 1. Values above 1 are divided by 0.8.

Octiles marked with shading.

*Dataset 2 – Revised Kauffman dataset*

Vector([0.04, .10, .10, .15, .15, .20, .30, .30, .30, .30, .40, .50, .60, .60, .60, .65, .65, .65, .65, .65, .65, .65, .70, .70, .70, .70, .75, .75, .75, .75, .75, .75, .80, .80, .85, .90, .90, .90, .90, .90, .90, .90, .90, .90, .90, .90, .90, .99, .99, .99, 1.3125, 1.375, 1.375, 1.375, 1.375, 1.375, 1.5, 1.5, 1.55, 1.5625, 1.5625, 1.625, 1.625, 1.625, 1.625, 1.6875, 1.6875, 1.6875, 1.6875, 1.6875, 1.6875, 1.75, 1.875, 1.875, 1.875, 2.0, 2.0, 2.125, 2.125, 2.125, 2.125, 2.125, 2.25, 2.625, 2.625, 2.75, 2.75, 2.75, 2.75, 2.875, 2.875, 2.875, 3.25, 3.75, 4, 4, 4, 4.75, 7.5, 10], datatype = float) Mean average = 1.555725

| | **1** | **2** | **3** | **4** | **5** | **6** | **7** | **8** |
|---|---|---|---|---|---|---|---|---|
| **Octile mean** | 0.24 | 0.65 | 0.79 | 0.92 | 1.46 | 1.72 | 2.30 | 4.26 |
| | **1** | **2** | **3** | **4** | | | | |
| **Quartile mean** | 0.452 | 0.857 | 1.595 | 3.32 | | | | |

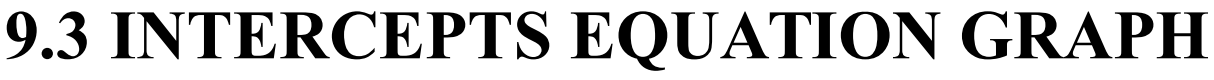

## 9.3 INTERCEPTS EQUATION GRAPH

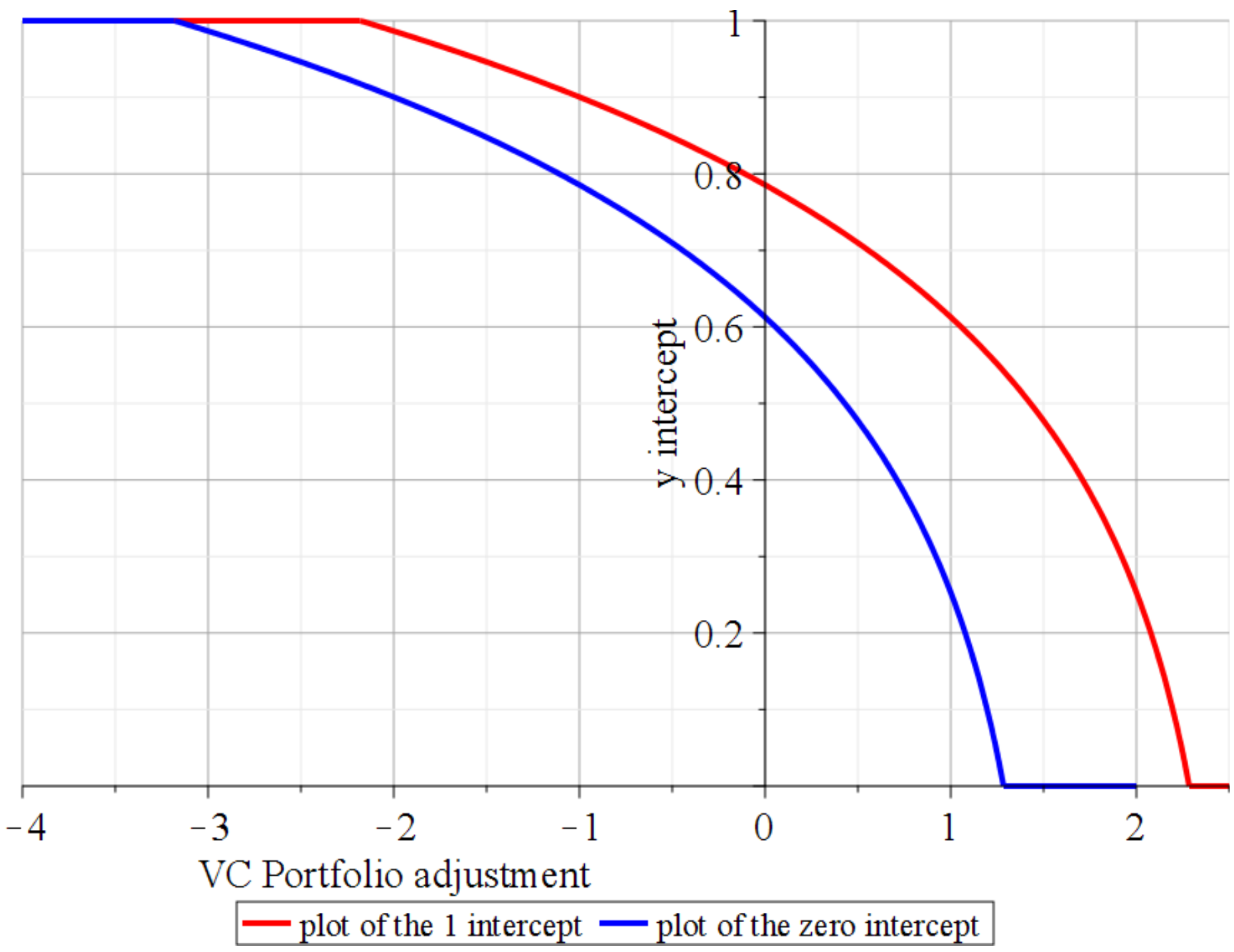

# 10 GLOSSARY

AIG – American International Group. A global company providing products to commercial, institutional and individual customers. They also provide mortgage and credit default swap (CDS) contracts.

BACPA – Bankruptcy Abuse Prevention and Consumer Protection Act of 2005. For these purposes, BACPA strengthened the rights of derivative holders to collect immediately.

Basel accords – There are three sets of banking regulations set by the Basel Committee on Bank Supervision. These are known as Basel I, Basel II and Basel III.

Call – In options trading, a call contract gives the holder the right to buy an asset at a pre-negotiated price for some time period. For a EDCS, it signals the side of the EDCS that collects the assets of an investment in return for payment of the value to the purchaser, and collects equity at exit of the investment.

CDS – Credit Default Swap. The purchaser makes premium payments to the underwriter and the contract insures a loan on some asset, typically a real estate loan. If the borrower defaults on the loan, then the purchaser is paid the face value of the contract, and transfers the asset to the underwriter.

DIN – Default Insurance Note. This is the most generic name for the type of contracts being used. The DIN is a superset of CDS, EDS, EDCS, and any other contract which exchanges equity in something for a payoff. However, these types of contracts are not technically insurance, although they can function as such in some situations. You may see ‘DIN’ appear in some of the variable names for equations.

EDS – Equity Default Swap. A proposed derivative that covers loans made by venture capitalists as investments, defined in this paper. It functions similarly to a CDS.

EDCS – Equity Default Clawback Swap. This improvement on the EDS that covers loans removes the perverse incentive to create losses. (The arsonist motive.) This EDCS contract is they key to enabling this new type of banking to function.

FRED – Federal Reserve Economic Data.

Haircut – A reduction in the stated value of an asset.

IPO – Initial Public Offering.

IP – Intellectual property. Patents, trademarks and copyrights.

M&A – Merger and Acquisition.

MOC – Multiple of Original Capital. Some amount of money is put into the bank that is its capital. This amount is enlarged by the Basel accords rules into the complete Tier 1 and Tier 2 capital that is used by the bank as reserves. The total outstanding investments divided by the original capital placed in bank Tier 1 reserves is the MOC. See figure 6.

Put – In options trading, a put is a contract that buys the right to force a buyer for your asset to pay a pre-negotiated price. For a EDCS, it is the right to collect the payoff amount of the and turn over the equity of an insured investment.

TAM – Total Available Market.

VBU – Venture-Bank Utility. This is a proposed new entity that handles the banking operations for a set of venture-banks. See figure 5.